\documentclass[aps, prd, twocolumn, nofootinbib, floatfix, superscriptaddress]{revtex4-1}
\usepackage{subfigure}
\usepackage{graphicx}
\usepackage{dcolumn}
\usepackage{epsfig}
\usepackage{amsmath}
\usepackage{amsfonts}
\usepackage{graphicx}
\usepackage{amssymb}
\usepackage{amssymb,amsmath,amsfonts}
\usepackage{xfrac}
\usepackage{color}
\usepackage{xcolor}
\usepackage{tikz}

\usepackage{tcolorbox}
\usepackage{hyperref}
\hypersetup{colorlinks, citecolor=bluscuro, linkcolor=black, urlcolor=bluscuro}
\definecolor{rossos}{cmyk}{0,1,1,0.55}
\definecolor{bluscuro}{rgb}{0.15, 0.2, .85}
\definecolor{bluchiaro}{cmyk}{1,.3,0.,0.1}

\newcommand{\be}{\begin{equation}}
\newcommand{\ee}{\end{equation}}
\newcommand{\bea}{\begin{eqnarray}}
\newcommand{\eea}{\end{eqnarray}}
\newcommand{\beq}{\begin{equation}}
\newcommand{\eeq}{\end{equation}}

\def\d{{\rm d}}

\newcommand{\mr}{\mathrm{r}}
\newcommand{\mt}{\mathrm{t}}
\newcommand{\Eq}[1]{Eq.~(\ref{#1})}
\newcommand{\Eqs}[1]{Eqs.~(\ref{#1})}
\newcommand{\Fig}[1]{Fig.~{\ref{#1}}}

\newcommand{\App}[1]{Appendix~\ref{#1}}

\newcommand{\Sec}[1]{Sec.~\ref{#1}}

\begin{document}

\title{ Primordial black hole formation for an anisotropic perfect fluid: \\ Initial conditions and estimation of the threshold }

\author{ Ilia Musco }
\email{ilia.musco@uniroma1.it}
\affiliation{Dipartimento di Fisica, Sapienza Università di Roma, Piazzale Aldo Moro 5, 00185, Roma, Italy}
\affiliation{INFN, Sezione di Roma, Piazzale Aldo Moro 2, 00185, Roma, Italy}

\author{Theodoros Papanikolaou}
\email{papaniko@noa.gr}
\affiliation{National Observatory of Athens, Lofos Nymfon, 11852 Athens, Greece}
\affiliation{Laboratoire Astroparticule et Cosmologie, CNRS Universit\'e de Paris, 75013 Paris, France }

\begin{abstract} 
\noindent
This work investigates the formation of primordial black holes within a radiation fluid with an anisotropic pressure. 
We focus our attention on the initial conditions describing cosmological perturbations in the super horizon 
regime, using a covariant form of the equation of state in terms of pressure and energy density gradients. The effect of 
the anisotropy  is to modify the initial shape of the cosmological perturbations with respect to the isotropic case. Using 
the dependence of the threshold $\delta_\mathrm{c}$ for primordial black holes with respect to the shape of cosmological 
perturbations, we estimate here how the threshold is varying with respect to the amplitude of the anisotropy. If this 
variation is large enough it could lead to a significant variation of the abundance of PBHs.
\end{abstract}

\maketitle


\section{Introduction}
\label{intro}
About 50 years ago it was already being argued that Primordial Black Holes (PBHs) might form during the radiation 
dominated era of the early Universe by gravitational collapse of sufficiently large-amplitude cosmological 
perturbations~\cite{Zeldovich:1967lct,Hawking:1971ei,Carr:1974nx} (see Refs.~\cite{Sasaki:2018dmp,Green:2020jor} for
recent reviews). This idea has recently received a lot of attention when it has been realized that PBHs could constitute a 
significant fraction of the dark matter in the Universe, see Ref.~\cite{Carr:2020gox} for a review of the current constraints 
onthe PBH abundances. This scenario is compatible with the gravitational waves detected during the O1/O2 and O3 
observational runs~\cite{LIGOScientific:2018mvr, LIGOScientific:2020stg,LIGOScientific:2020zkf,LIGOScientific:2020iuh} 
of the LIGO/Virgo Collaboration, and has motivated several studies concerrning the primordial origin of these events~\cite{Sasaki:2016jop,Bird:2016dcv,Clesse:2016vqa, Ali-Haimoud:2017rtz,Raidal:2018bbj,Hutsi:2019hlw,Vaskonen:2019jpv, Gow:2019pok, DeLuca:2020fpg, DeLuca:2020qqa, Clesse:2020ghq, Hall:2020daa, Jedamzik:2020ypm, Jedamzik:2020omx, DeLuca:2020sae, DeLuca:2020jug}. 
In particular, the GWTC-2 catalog is found to be compatible with the primordial scenario \cite{Wong:2020yig} and a possible 
detection of a stochastic gravitational wave background by the NANOGrav collaboration~\cite{NANOGrav:2020bcs} could be 
ascribed to PBHs~\cite{Vaskonen:2020lbd, DeLuca:2020agl, Kohri:2020qqd, Domenech:2020ers, Sugiyama:2020roc,Inomata:2020xad}.
 
Despite some pioneering numerical studies~\cite{1978SvA....22..129N,1979ApJ...232..670B,1979sgrr.work..173N}, it has only 
recently become possible to fully understand the mechanism of PBH formation with detailed spherically symmetric numerical 
simulations~\cite{Jedamzik:1999am,Shibata:1999zs,Hawke:2002rf,Musco:2004ak,Polnarev:2006aa,Musco:2008hv,Musco:2012au}, 
showing that cosmological perturbations can collapse to PBHs if their amplitude $\delta$, measured at horizon crossing, 
is larger than a certain threshold value $\delta_c$. This quantity was initially estimated with a simplified Jeans length 
argument in Newtonian gravity~\cite{Harada:2013epa}, obtaining $\delta_c \sim c_s^2$, where $c_s^2=1/3$ is the sound 
speed of the cosmological radiation fluid measured in units of the speed of light. 

This estimation was then refined generalizing the Jeans length argument within the theory of General Relativity, which 
gives $\delta_c\simeq0.4$ for a radiation dominated Universe~\cite{Harada:2013epa}. This analytical computation however 
does not take into account the non linear effects of pressure gradients, related to the particular shape of the collapsing 
cosmological perturbation, which require full numerical relativistic simulations. A recent detailed study has shown a clear
relation between the value of the threshold $\delta_c$ and the initial curvature (or energy density) profile, with 
$0.4\leq \delta_c \leq 2/3$, where the shape is identified by a single parameter~\cite{Musco:2018rwt,Escriva:2019phb}. 
This range is reduced to $0.4\leq \delta_c \lesssim 0.6$ when the initial perturbations are computed from the primordial 
power spectrum of cosmological perturbations~\cite{Musco:2020jjb}, because of the smoothing associated with very large 
peaks. 

All of these spherically symmetric numerical simulations have considered the radiation Universe as isotropic, an 
approximation which is well justified in the context of peak theory, where rare large peaks which collapse to form PBHs 
are expected to be quasi spherical~\cite{Bardeen:1985tr}. However, it is very interesting to go beyond such assumptions 
and have a more realistic treatment of the gravitational collapse of cosmological perturbations. 

Regarding the spherical symmetry hypothesis, there were some early studies going beyond this and adopting the 
“pancake” collapse ~\cite{1965ApJ...142.1431L,1970Ap......6..320D, 1970A&A.....5...84Z, 1980PhLB...97..383K} as well 
as some recent ones focusing on a non-spherical collapse to form PBHs in a matter dominated 
Universe~\cite{Harada:2015ewt} and on the ellipsoidal collapse to form PBHs ~\cite{Kuhnel:2016exn}. 

To the best of our knowledge there has not yet been any systematic treatment of gravitational collapse of cosmological 
perturbations for anisotropic fluids. In general, one expects that anisotropies will arises in the presence of scalar fields 
and multifluids having, in spherical symmetry, a radial pressure component which is different from the tangential 
one~\cite{1982NCimB..69..145L}.  Substantial progress have been made in the analysis of anisotropic relativistic star 
solutions, using both analytical~\cite{1974ApJ...188..657B,1980PhRvD..22..807L,Bayin:1982vw,Mak:2001eb,Dev:2004ss,Herrera:2004xc,Veneroni:2018gfm} and numerical techniques~\cite{Doneva:2012rd,Biswas:2019gkw}. 

More recently a covariant formulation of the equation of state has been proposed with the study of equilibrium models of 
anisotropic stars as ultracompact objects behaving as black-holes~\cite{Raposo:2018rjn}. Inspired by this recent work, we 
study here the anisotropic formulation of the initial conditions for the collapse of cosmological perturbations, estimating the 
effect of the anisotropy on the threshold $\delta_\mathrm{c}$ for PBH formation. 

Following this introduction, in \Sec{Misner-Sharp} we recap the system of Einstein plus hydrodynamic equations for an 
anisotropic perfect fluid, introducing then in \Sec{EoS} the covariant formulation of the equation of state in terms of pressure 
and energy density gradients. In \Sec{math} we describe the gradient expansion approximation to set up the mathematical 
description of the initial conditions for this system of equations computed explicitly in \Sec{initial} for the different choice of 
the equation of state described in \Sec{EoS}. With this, in \Sec{results} we estimate the corresponding threshold for PBH 
formation, assuming that it varies with the shape of the initial energy density perturbation profile in the same way as for the 
isotropic case. Finally, in \Sec{conclusions} we summarize our results drawing some conclusions and discussing the future 
perspectives for this work. Throughout we use $c = G = 1$.


\section{Misner-Sharp equations for anisotropic fluids}
\label{Misner-Sharp}
In the following, we are going to revise, assuming spherical symmetry, the Einstein and hydrodynamic equations for an 
anisotropic perfect fluid. Using the cosmic time slicing the metric of space time can be written in a diagonal form as 
\beq\label{MS-Metric}
\mathrm{d}s^2=-A^2(t,r)\mathrm{d}t^2+B^2(t,r)\mathrm{d}r^2+R^2(t,r)\mathrm{d}\Omega^2
\eeq
where $r$ is the radial comoving coordinate, $t$ the cosmic time coordinate and $\mathrm{d}\Omega$ the solid line 
infinitesimal element of a unit $2$-sphere, i.e. $\mathrm{d}\Omega^2 = \mathrm{d}\theta^2 + \sin^2\theta\mathrm{d}\phi^2$. 
In this slicing there are three non zero components of the metric, which are functions of $r$ and $t$: the lapse function 
$A(r,t)$, the function $B(r,t)$ related to the spatial curvature of space time and the areal radius $R(r,t)$. 
The metric \eqref{MS-Metric} reduces to the Friedmann-Lema\^itre-Robertson-Walker (FLRW) form when the Universe is homogeneous and isotropic, with $A=1$ 
(normalization choice), $B=a(t)/\sqrt{1 - Kr^2}$ and $R=a(r)r$, with $a(t)$ being the scale factor, and $K=0,\pm1$ 
measuring the spatial curvature of the homogeneous Universe. 
 
In the Misner-Sharp formulation of the Einstein plus hydro equations ~\cite{Misner:1964je}, it is useful to introduce the 
differential operators $D_t$ and $D_r$ defined as
\beq\label{D_r and D_t in Misner-Sharp}
 D_t\equiv \left. \frac{1}{A}\frac{\partial}{\partial t}\right\vert_r \mathrm{\quad and \;\;\;\;} D_r\equiv \left.\frac{1}{B}\frac{\partial}{\partial r}\right\vert_t,
\eeq     
which allow one to define and compute the derivatives of the areal radius $R$ with respect to proper time and proper 
distance respectively. This introduces two auxiliary quantities, 
\beq\label{U and Gamma}
U\equiv D_tR \mathrm{\;\;\;\;and\;\;\;\;} \Gamma\equiv D_r R,
\eeq
where $U$ is the radial component of the four-velocity in the ``Eulerian" (non comoving) frame and $\Gamma$ is the so 
called generalized Lorentz factor introduced by Misner~\cite{Misner:1964je}. In the homogeneous and isotropic FLRW Universe,  
according to the Hubble law we have $U=H(t)R(t,r)$, and \mbox{$\Gamma^2 = 1-Kr^2$}, where $H(t)=\dot{a}(t)/a(t)$ is the 
Hubble parameter and $\dot{a}\equiv \partial a / \partial t$. 

The quantities $U$ and $\Gamma$ are related through the Misner-Sharp mass $M(r,t)$, defined within spherical symmetry 
as ~\cite{Misner:1964je, Hayward:1994bu} 
\beq\label{Misner Sharp Mass}
M(t,r)\equiv \frac{R(t,r)}{2}\left[1-\nabla_\mathrm{\mu}R(t,r)\nabla^\mathrm{\mu}R(t,r)\right], 
\eeq
and from the above definition one can get the constraint equation 
\beq\label{constraint equation}
\Gamma^2=1+U^2-\frac{2M}{R}
\eeq
obtained by integrating the 00-component of the Einstein equations. 

The stress-energy tensor for an anisotropic perfect fluid can be written in a covariant form~\cite{Raposo:2018rjn} as
\beq\label{Stress Energy Tensor}
T_{\mathrm{\mu\nu}}= \rho u_\mathrm{\mu} u_\mathrm{\nu} +p_\mr k_\mathrm{\mu}k_\mathrm{\nu} +p_\mt \Pi_\mathrm{\mu\nu},
\eeq
where $p_r$ and $p_t$ are the radial and the tangential pressure respectively, $u_\mathrm{\mu}$ is the fluid four-velocity 
and $k_\mathrm{\mu}$ is a unit spacelike vector orthogonal to $u_\mathrm{\mu}$, i.e. \mbox{ $u_\mathrm{\mu}
u^{\mathrm{\mu}}=-1$}, \mbox{ $k_\mathrm{\mu}k^\mathrm{\mu}=1$} and $u^{\mathrm{\mu}}k_\mathrm{\mu}=0$. 
$\Pi_\mathrm{\mu\nu} = g_\mathrm{\mu\nu}+u_\mathrm{\mu}u_\mathrm{\nu}-k_\mathrm{\mu}k_\mathrm{\nu}$ 
is a projection tensor onto a two surface orthogonal to $u^\mathrm{\mu}$ and $k^\mathrm{\mu}$. Working in the comoving 
frame of the fluid one obtains that $u_\mathrm{\mu} = (-A,0,0,0)$ and $k_\mathrm{\mu} = (0,B,0,0)$. In the limit of 
$p_\mr = p_\mathrm{t}$ the stress energy tensor reduces to the standard isotropic form.

Considering now the Einstein field equations and the conservation of the stress energy tensor, respectively given by
\beq 
G^{\mathrm{\mu}\mathrm{\nu}}=8\pi T^{\mathrm{\mu}\mathrm{\nu}}  \quad\quad  \nabla_\mathrm{\mu}T^\mathrm{\mu\nu}=0 \,,
\eeq
where $G^{\mathrm{\mu}\mathrm{\nu}}$ is the Einstein tensor, the Misner-Sharp hydrodynamic 
equations~\cite{Misner:1964je, May:1966zz} for an anisotropic spherically symmetric fluid are given by:
\beq \label{MS-Equations}
\begin{aligned}
& D_tU = - \frac{\Gamma}{\rho+p_\mathrm{r}} \left[ D_r p_\mathrm{r} \!+\!  \frac{2\Gamma}{R} \left( p_\mathrm{r}\!-\!p_\mathrm{t} \right) \right] \! - \! \frac{M}{R^2} \! - \! 4\pi Rp_\mathrm{r}   \\
& \frac{D_t\rho_0}{\rho_0} = - \frac{1}{R^2\Gamma}D_r\left(R^2U\right) \\
& \frac{D_t\rho}{\rho+p_\mathrm{r}} = \frac{D_t\rho_0}{\rho_0} + \frac{2U}{R} \frac{p_\mathrm{r}-p_\mathrm{t}}{\rho+p_\mathrm{r}}  \\ 
& \frac{D_rA}{A} = -\frac{1}{\rho+p_\mathrm{r}} \left[ D_r p_\mathrm{r} + \frac{2\Gamma}{R} \left(p_\mathrm{r}-p_\mathrm{t}\right) \right] \\ 
& D_rM = 4\pi R^2\Gamma\rho  \\
& D_tM=-4\pi R^2 Up_\mathrm{r}  \\ 
& D_t\Gamma = - \frac{U}{\rho+p_\mathrm{r}} \left[ D_rp_\mathrm{r} + \frac{2\Gamma}{R} \left(p_\mathrm{r}-p_\mathrm{t}\right) \right]\,,
\end{aligned}
\eeq
where one can appreciate the additional terms appearing in the equations when $p_\mr \neq p_\mathrm{t}$. 
\noindent


\section{Equation of state for anisotropic pressure}
\label{EoS}
We introduce here a covariant formulation of the equation of state for an anisotropic perfect fluid, where the difference
between the radial and tangential pressures is measured in terms of pressure or energy density gradients. In particular, 
following~\cite{Raposo:2018rjn, Bowers:1974tgi} the difference $p_\mt- p_\mr$ can be expressed, up to a certain degree 
of arbitrariness, in a covariant form as 
\begin{eqnarray}
p_\mt & = & p_\mr + \lambda g(r,t) k^\mu \nabla_\mu p_\mr  \label{D_rp_r} \\ 
& & \quad \quad \quad \textrm{or} \nonumber \\
p_\mt & = & p_\mr + \lambda g(r,t) k^\mu \nabla_\mu \rho  \label{D_rrho},
\end{eqnarray}
where $g(r,t)$ is a generic function of $r$ and $t$ while $\lambda$ is a parameter tuning the 
level of the anisotropy. 

Equations \eqref{D_rp_r} and \eqref{D_rrho} are two possible ways to express 
in covariant form the difference $(p_\mathrm{r} - p_\mathrm{t})$, without specifying explicitly the underlying microphysics. 
The most general way to do it can be found in Appendix A of~\cite{Raposo:2018rjn}. In general the parametrization of the EoS 
depends on the microphysics of the fluid, in particular on the interactions between the fluid 
particles~\cite{Bowers:1973zza,Bowers:1974tgi}.

Because we are considering a radiation dominated medium, it looks reasonable to assume the conservation of the trace 
of the stress-energy tensor, i.e. $T^\mathrm{\mu}_\mathrm{\mu}=0$, giving an additional constraint 
relation\footnote{ For a relativistic fluid, $E\gg m$ and the fluid particles can be considered as massless with 
the norm of the four-momentum being very close zero, i.e. $k^\alpha k_\alpha \simeq 0$, having as a consequence the 
stress-energy tensor being traceless~\cite{Ellis:1971pg}.}
\beq\label{trace constraint}
\rho - p_\mathrm{r} - 2p_\mathrm{t} = 0
\eeq
which, together with \eqref{D_rp_r} or \eqref{D_rrho}, gives closure of the system of equations to be solve.

Looking at the form of the Minser-Sharp equations given by \eqref{MS-Equations} we need to make sure that the behavior 
at $R=0$ is regular~\cite{Bowers:1974tgi}, implying
\beq\label{boundary condition}
\lim_{R\rightarrow 0} \frac{p_\mathrm{r}-p_\mathrm{t}}{R} = 0\,.
\eeq
 This can be obtained choosing $g(r,t)=R(r,t)$ which compensates the $1/R$ term appearing in the anisotropic terms of the 
 Misner-Sharp equations, keeping the parameter $\lambda$ dimensionless, without introducing an additional characteristic 
 scale into the problem. In this case, using $k^\mu \nabla_\mu = D_r$combined with  \Eq{D_rp_r} and \Eq{trace constraint}, 
 the equations of state (EoS) for $p_\mr$ and $p_\mt$ read as
\beq\label{eq:p_r+p_t - f= R-D_rp_r} 
p_\mathrm{r} =  \frac{1}{3}\left[\rho-2\lambda R D_rp_\mr \right] \quad 
p_\mathrm{t} = \frac{1}{3}\left[\rho+\lambda R D_rp_\mr \right] ,
\eeq
while when we combine \Eq{D_rrho} with \Eq{trace constraint}, the equations of state (EoS) are given by 
\beq \label{eq:p_r+p_t - f= R-D_rrho} 
p_\mathrm{r}  =  \frac{1}{3}\left[\rho-2\lambda R D_r\rho \right]\quad p_\mathrm{t} = 
\frac{1}{3}\left[\rho+\lambda RD_r\rho\right]. 
\eeq
Another interesting possibility is to choose \mbox{$g(r,t)=\rho^n(r,t)$}, where $n$ is an integer. In that case, the anisotropy 
parameter $\lambda$ is not dimensionless, but the equations of state for $p_\mr$ and $p_\mt$ depend only on local 
thermodynamic quantities of the comoving fluid element, a key difference with respect to the previous case where the 
choice of $g(r,t)=R(r,t)$ makes the EoS fully non local. Using this second choice for $g(r,t)$, if the EoS is given by 
\Eq{D_rp_r} one obtains that 
\beq\label{eq:p_r+p_t - f= rho^n-D_rp_r}
p_\mathrm{r}  = \frac{1}{3}\left[\rho-2\lambda \rho^n D_rp_\mr \right]\quad p_\mathrm{t} = 
\frac{1}{3}\left[\rho+\lambda \rho^n D_rp_\mr \right]
\eeq
while when the EoS is given by \Eq{D_rrho} one has
\beq \label{eq:p_r+p_t - f= rho^n-D_rrho}
p_\mathrm{r}  =  \frac{1}{3}\left[\rho-2\lambda \rho^nD_r\rho \right]\quad p_\mathrm{t} = 
\frac{1}{3}\left[\rho+\lambda \rho^n D_r\rho\right].
\eeq
As one can see, when \mbox{$\lambda=0$} the fluid is isotropic and these expressions reduce to the standard EoS 
$p_\mathrm{r}=p_\mathrm{t}=\rho/3$ for an isotropic relativistic perfect fluid.
 
\noindent
 
\section{Initial conditions: Mathematical formulation}
\label{math}
\subsection{The curvature profile}
PBHs are formed from the collapse of nonlinear cosmological perturbations after they reenter the cosmological horizon. 
Following the standard result for large and rare peaks we assume spherical symmetry on superhorizon scales, 
where the local region of the Universe is characterized by an asymptotic solution ($t\rightarrow 0$) of Einstein’s 
equations~\cite{Lifshitz:1963ps}. In this regime the asymptotic metric can be written as
\beq\label{FRLW metric-K(r) form}
\mathrm{d}s^2 = -\mathrm{d}t^2 + a^2(t)\left[\frac{\mathrm{d}r^2}{1-K(r)r^2}+r^2\mathrm{d}\Omega^2\right]\,.
\eeq
where $K(r)$ is the initial curvature profile for adiabatic perturbations, written as a perturbation of the 3-spatial metric,  
which is time independent on superhorizon scales. 

An alternative way to specify the curvature profile for adiabatic cosmological perturbations is the function $\zeta(\tilde{r})$, 
perturbing the scale factor, with the asymptotic metric given by
\beq
\mathrm{d}s^2 = -\mathrm{d}t^2 + a^2(t)e^{2\zeta(\tilde{r})} \left[ \mathrm{d}\tilde{r}^2 + \tilde{r}^2\mathrm{d}\Omega^2\right ]\,,
\eeq
where $r=\tilde{r}e^{\zeta(\tilde{r})}$. 

In the following we are going to describe the initial conditions only in terms of $K(r)$, which allows a simpler mathematical 
description, although one can always express them with $\zeta(\tilde{r})$, by making a coordinate 
transformation~\cite{Musco:2018rwt}. This will be useful to connect the initial conditions to the power spectrum of 
cosmological perturbations~\cite{Musco:2020jjb}. 

\subsection{Gradient expansion approximation}
Although the initial amplitude of the curvature profile is non linear for perturbations giving rise to PBH formation, the 
corresponding hydrodynamic perturbations, in energy density and velocity, are time dependent and vanish asymptotically
going backwards in time (as $t\rightarrow 0$). These can be treated as small perturbations on superhorizon scales, 
when the perturbed regions are still expanding, parametrized by a small parameter 
$\epsilon$ defined as the ratio between the Hubble radius $H^{-1}$ and a characteristic scale $L$ (to be defined later),
\beq\label{epsilon definition}
\epsilon(t) \equiv \frac{H^{-1}}{L}  \ll 1.
\eeq

In the superhorizon regime, pure growing modes are of $O(\epsilon^2)$ in the first non zero term of the 
expansion~\cite{Lyth:2004gb,Musco:2018rwt}. This approach is known in the literature as the long 
wavelength~\cite{Shibata_1999}, gradient expansion~\cite{Salopek:1990jq} or separate universe 
approach~\cite{Wands:2000dp,Lyth:2004gb} and reproduces the time 
evolution of the linear perturbation theory. The hydrodynamic variables $\rho$, $U$, $p_\mr$, $p_\mt$ and $M$, and the
metric ones $A$, $B$ and $R$, can be expanded as~\cite{Polnarev:2006aa}
\begin{equation}\label{Perturbed Variables}
\begin{split}
\rho & = \rho_\mathrm{b}(t)\left[1 + \epsilon^2 \tilde{\rho}(r,t)\right] \\
p_\mathrm{r} & = \frac{\rho_\mathrm{b}(t)}{3}\left[1+\epsilon^2\tilde{p}_\mathrm{r}(r,t)\right] \\ 
p_\mathrm{t} & = \frac{\rho_\mathrm{b}(t)}{3}\left[1+\epsilon^2\tilde{p}_\mathrm{t}(r,t)\right] \\ 
U & = H(t)R\left[1+\epsilon^2\tilde{U}(r,t)\right]\\
M & = \frac{4\pi}{3}\rho_\mathrm{b}(t)R^3\left[1+\epsilon^2\tilde{M}(r,t)\right] \\
 A & = 1 + \epsilon^2 \tilde{A}(r,t) \\
B & = \frac{R^\prime}{\sqrt{1-K(r)r^2}}\left[1+\epsilon^2\tilde{B}(r,t)\right] \\
R & = a(t)r\left[1+\epsilon^2\tilde{R}(r,t)\right] \,.
\end{split}
\end{equation}
where one should note that the multiplicative terms outside the parentheses do not always correspond to the background
values. Looking at the velocity $U$, for example, the perturbation of the Hubble parameter, described by $\tilde{U}(r,t)$, 
is separated with respect to the perturbation of the areal radius given by $\tilde{R}(r,t)$. 

\subsection{The perturbation amplitude}
\label{threshold}
Before perturbing the Misner-Sharp equations in the next section, we introduce at this stage the definition of the 
perturbation amplitude, consistent with the criterion to find when a cosmological perturbation is able to form 
a PBH. This depends on the amplitude measured at the peak of the compaction function~\cite{Shibata:1999zs} 
defined as
\be
\label{C definition}
\mathcal{C} \equiv 2\frac{\delta M(r,t)}{R(r,t)} \,,
\ee
where $\delta M(r,t)$ is the difference between the Misner-Sharp mass within a sphere of radius $R(r,t)$, and the 
background mass \mbox{$M_b(r,t)=4\pi \rho_b(r,t)R^3(r,t)/3$} within the same areal radius, but calculated with 
respect to a spatially flat FLRW metric. As shown in \cite{Musco:2018rwt}, according to this criterion, the comoving length 
scale of the perturbation should be identified with $r=r_m$, where the compaction function reaches 
its maximum \mbox{(i.e. \mbox{$\mathcal{C}'(r_m) = 0$})} with the perturbation scale measured with respect the 
background, i.e. $L \equiv a r_m$ and
\be
\epsilon = \frac{1}{aHr_m} \,.
\ee

The perturbation amplitude is defined as the mass excess of the energy density within the scale $r_m$, measured at the
cosmological horizon crossing time $t_H$, defined when $\epsilon=1$ ($aHr_m=1$). Although in this regime the gradient
expansion approximation is not very accurate, and the horizon crossing defined in this way is only a linear extrapolation, 
this provides a well defined criterion to measure consistently the amplitude of different perturbations, understanding how 
the threshold is varying because of the different initial curvature profiles (see~\cite{Musco:2018rwt} for more details). 

The amplitude of the perturbation measured at $t_H$, which we refer to just as $\delta \equiv \delta(r_m,t_H)$, is given by 
the excess of mass averaged over a spherical volume of radius $R_m$, defined as
\be
\label{delta}
\delta \equiv \frac{4\pi}{V_{R_m}} \int_0^{R_m}   \frac{\delta\rho}{\rho_b} \,R^2 \d R\,  =
\frac{3}{r_m^3} \int_0^{r_m} \frac{\delta\rho}{\rho_b} \, r^2 \d r  \,,
\ee
where $V_{R_m} = {4\pi}R_m^3/3$. The second equality is obtained by neglecting the higher order terms in $\epsilon$, 
approximating \mbox{$R_m \simeq a(t)r_m$}, which allows one to simply integrate over the comoving volume of radius $r_m$.

\section{Initial conditions: Anisotropic quasi-homogeneous solution}
\label{initial}
We are now ready to perform the perturbative analysis, computing the initial conditions as functions of the curvature profile 
$K(r)$. Introducing \eqref{Perturbed Variables} into the the Misner-Sharp equations given by 
\eqref{MS-Equations} one gets the following set of differential equations:
\beq\label{eq:metric+hydrodynamic perturbations}
\begin{aligned}
2\tilde{R}+\frac{\partial\tilde{R}}{\partial N }  & =  \tilde{A}+\tilde{U} \\
2\tilde{B}+\frac{\partial\tilde{B}}{\partial N} & =  -r \tilde{A}^\prime \\
\tilde{A}^\prime & =  - \frac{1}{4}\left[\tilde{p}^\prime_\mathrm{r}+\frac{2}{r} \left(\tilde{p}_\mathrm{r}-\tilde{p}_\mathrm{t}\right)\right] \\
\tilde{\rho} & = \frac{1}{3r^2}\left(r^3\tilde{M}\right)^\prime \\
\tilde{M}+\frac{\partial\tilde{M}}{\partial N} & = -4\tilde{U}-4\tilde{A}-\tilde{p}_\mathrm{r} \\
\tilde{U} & = \frac{1}{2}\left[\tilde{M}-K(r)r_\mathrm{m}^2\right],
\end{aligned}
\eeq
where $N\equiv\ln(a/a_\mathrm{i})$ is measuring the number of e-foldings, and $a_\mathrm{i}$  is the scale factor 
computed at an initial time $t_\mathrm{i}$. In the following, we solve this set of equations using the EoS described 
earlier in \Sec{EoS}. 

\subsection{Equation of state with $g(r,t)=R(r,t)$}\label{f=R}
At a first glance, the Misner-Sharp equations obtained in \eqref{MS-Equations} could have a non regular behavior in 
the center ($R=0$) because of the anisotropic corrections given by the two terms: 
\[ 2\frac{U}{R} \left( p_\mathrm{r}-p_\mathrm{t} \right) \quad \textrm{and} \quad 2\frac{\Gamma}{R} \left( p_\mathrm{r}-p_\mathrm{t} \right)  \,.\]
The first one is naturally cured by the behavior of  $U\sim HR$, as specified in \eqref{Perturbed Variables}, while the 
second term, having $\Gamma(0) = 1$, requires a careful choice of the energy density profile, which will determine the
difference $\left( p_\mathrm{r}-p_\mathrm{t} \right)$. However this problem can be avoided with a careful choice of $g(r,t)$: 
in particular choosing $g(r,t) = R(r,t)$ is both canceling the possible divergence and making $\lambda$ a naturally scale 
independent parameter, having in this way a scale-free problem as in the isotropic case. 

This choice looks mathematically elegant and simple, but it has the drawback of introducing into the EoS a non local quantity, 
namely $R(r,t)$. Although it looks to be ad-hoc, it is useful to analyze such a case as a simple toy model in order to study 
the structure of the solution of the system of equations \eqref{eq:metric+hydrodynamic perturbations}. 

In this case, the explicit equations for the perturbation of the radial pressure $\tilde{p}_\mr$ and the lapse 
perturbation $\tilde{A}$ are given by
\beq\label{p_tilde-rho_tilde}
\tilde{p}_\mathrm{r}-\tilde{\rho}=-\frac{2\lambda}{3} rf(r) \,,
\eeq
\beq\label{A_tilde constraint - Drpr}
\tilde{A}^\prime = - \frac{1}{4}\left[ \tilde{p}^\prime_\mathrm{r} - 2 \lambda f(r) \right]\,,
\eeq
where
\begin{equation} \label{f_j:f=R}
f(r) = (2j+1) \sqrt{1-K(r)r^2} \cdot \left\{ 
\begin{split}
\tilde{p}^{\prime}_\mathrm{r} \quad   \textrm{if \ $j=0$} \\ 
\tilde{\rho}^{\prime} \quad  \textrm{if \ $j=1$}
\end{split} \right.
\end{equation}
The index $j$ allows to distinguishing between \Eq{eq:p_r+p_t - f= R-D_rp_r} where the EoS is expressed in terms of 
pressure gradients \mbox{($j=0$)} and \Eq{eq:p_r+p_t - f= R-D_rrho} when the EoS is expressed in terms of density gradients ($j=1$).

Inserting \Eqs{p_tilde-rho_tilde} and \eqref{A_tilde constraint - Drpr} into \eqref{eq:metric+hydrodynamic perturbations} one finds the explicit quasi-homogeneous solution of the initial perturbation profiles as a function of the curvature profile $K(r)$: 
\begin{eqnarray}\label{Perturbations}
\begin{split}
\tilde{\rho} & = \frac{2}{3}\frac{\left[r^3\mathcal{K}(r)\right]^\prime}{3r^2}r^2_\mathrm{m}  \\
\tilde{U} & =-\frac{1}{6}\mathcal{K}(r)r_\mathrm{m}^2 - \frac{\lambda}{2}\mathcal{F}(r)  \\ 
\tilde{M} &=\frac{2}{3}\mathcal{K}(r)r_\mathrm{m}^2 \\
\tilde{A} &  = - \frac{\tilde{\rho}}{4} +\frac{\lambda}{2}\frac{\left[r^3\mathcal{F}(r)\right]^\prime}{3r^2}  \\ 
\tilde{B} &  =  r \left[\frac{\tilde{\rho}}{8}  - \frac{\lambda}{4} \frac{\left[r^3\mathcal{F}(r)\right]^\prime}{3r^2}\right]^\prime 
\\
\tilde{R} & = -\frac{\tilde{\rho}}{8} + \frac{\tilde{U}}{2} + \frac{\lambda}{4}\frac{\left[r^3\mathcal{F}(r)\right]^\prime}{3r^2}  \\
\end{split}
\end{eqnarray}
where $\mathcal{K}(r)$ is an effective curvature profile 
\beq\label{Effective Curvature Profile-Drpr}
\mathcal{K}(r) \equiv K(r) - \frac{\lambda}{r^2_\mathrm{m}}\mathcal{F}(r)
\eeq
and 
\beq\label{F:f=R}
\mathcal{F}(r)=  \int_{\infty}^r f(r^\prime)\mathrm{d}r^\prime \,,
\eeq 
is sourcing the anisotropic modification of the quasi-homogeneous solution. In \App{app:The pressure and energy density gradient profiles}, we show how to 
compute explicitly the profile of $f(r)$, analyzing how it is varying with $\lambda$.

It is easy to see that, when $\lambda=0$, from \Eq{p_tilde-rho_tilde} and \eqref{A_tilde constraint - Drpr} we simply 
have $\tilde{p}= \tilde{\rho} = 4\tilde{A}$, canceling the two last terms of the differential equation for 
$\tilde{M}$ in \eqref{eq:metric+hydrodynamic perturbations}, and from \eqref{Perturbations} one is recovering the 
quasi-homogeneous solution for an isotropic radiation fluid, which has been derived in \cite{Polnarev:2006aa}, and 
more extensively discussed in \cite{Musco:2018rwt}. 

The effective curvature profile $\mathcal{K}(r)$ allows writing the anisotropic quasi-homogeneous solution in a form which 
is very similar to the isotropic case ($\lambda=0$). Following this strategy one can introduce effective energy density 
and velocity perturbations, 
 $\tilde{\rho}_\mathrm{eff}$ and $\tilde{U}_\mathrm{eff}$, defined as
\beq\label{rho_tilde_eff}
\tilde{\rho}_\mathrm{eff} = \tilde{\rho}-2\lambda\frac{\left[r^3\mathcal{F}(r)\right]^\prime}{3r^2}
\eeq
\beq\label{U__tilde_eff}
\tilde{U}_\mathrm{eff} = \tilde{U}+\frac{\lambda}{2}\mathcal{F}(r) = -\frac{1}{6}\mathcal{K}(r)r^2_\mathrm{m}
\eeq
where one can appreciate that $\tilde{U}_\mathrm{eff}$ expressed in terms of the effective curvature $\mathcal{K}(r)$ takes 
the same form as in the isotropic case. The effective energy density and velocity perturbations allow writing all of the other 
perturbed variables just as linear combinations of these two quantities
\begin{eqnarray}\label{Perturbations_eff}
\begin{split}
\tilde{M} & = - 4\tilde{U}_\mathrm{eff}  \\
\tilde{A} &  = - \frac{\tilde{\rho}_\mathrm{eff}}{4} \\
\tilde{B} &  = \frac{r}{8}\tilde{\rho}^\prime_\mathrm{eff} \\
\tilde{R} & = - \frac{\tilde{\rho}_\mathrm{eff}}{8} + \frac{\tilde{U}}{2} 
\end{split}
\end{eqnarray}
keeping the same functional form as the isotropic solution (see \cite{Musco:2018rwt} for more details).

\subsection{Equation of state with $g(r,t)=\rho^n(r,t)$}\label{f=rho^n}
An alternative choice for the equation of state is  \mbox{$g(r,t)=\rho^n(r,t)$} as suggested in~\cite{Raposo:2018rjn}, 
motivated by physical considerations based on a microphysical description of the matter. This makes the EoS for 
$p_\mr$ and $p_\mt$ just a function of local thermodynamic quantities. However because in \Eqs{MS-Equations} 
the anisotropic terms $(p_\mathrm{r}-p_\mathrm{t})$ are multiplied by $1/R$, one then needs to require that 
$\rho^n(D_r p_\mr)/R$ should vanish at least as $R\rightarrow 0$, in order to ensure a regular behaviour in the 
center ($R=0$).

In this case, the explicit equations to compute the perturbation of the radial pressure $\tilde{p}_\mr$ and the lapse perturbation $\tilde{A}$ become
\beq\label{p_tilde-rho_tilde - f=rho^n}
\tilde{p}_\mathrm{r}-\tilde{\rho} = - \frac{2\lambda}{3} \frac{\rho^n_\mathrm{b}(a)}{a}r f(r) 
\eeq
\beq\label{A_tilde constraint - f=rho^n}
\tilde{A}^\prime = - \frac{1}{4} \left[ \tilde{p}^\prime_\mathrm{r} - 2\lambda\frac{\rho^n_\mathrm{b}(a)}{a}f(r)\right]
\eeq
where  this time $f(r)$ is defined as 
\begin{equation} \label{f_j:f=rho^n}
f(r) = (2j+1) \frac{\sqrt{1-K(r)r^2}}{r} \cdot \left\{ 
\begin{split}
\tilde{p}^{\prime}_\mathrm{r} \quad   \textrm{if \ $j=0$} \\ 
\tilde{\rho}^{\prime} \quad   \textrm{if \ $j=1$}
\end{split} \right.
\end{equation}
As in the previous section, for $j=0$ the EoS is expressed in terms of pressure gradients, following now 
\Eq{eq:p_r+p_t - f= rho^n-D_rp_r}, while for $j=1$ it is expressed in terms of energy density gradients, corresponding to
\Eq{eq:p_r+p_t - f= rho^n-D_rrho}.

Inserting these expressions into \eqref{eq:metric+hydrodynamic perturbations} one obtains the following quasi-homogeneous 
solution 
\begin{eqnarray}\label{Perturbations-f=rho^n}
\begin{split}
\tilde{\rho} & = \frac{2}{3}\frac{\left[r^3\mathcal{K}(r)\right]^\prime}{3r^2}r^2_\mathrm{m} \\ 
\tilde{U} & = - \frac{1}{6} \mathcal{K}(r)r^2_\mathrm{m} - \frac{\lambda}{2}\Phi(a)\mathcal{F}(r) \\
\tilde{M} & = \frac{2}{3}\mathcal{K}(r)r^2_\mathrm{m}  \\
\tilde{A} & =  -\frac{1}{4}\tilde{\rho} + \frac{\lambda}{2} \frac{\rho^n_\mathrm{b}(a)}{ar_\mathrm{m}}  \frac{\left[r^3\mathcal{F}(r)\right]^\prime}{3r^2} \\
\tilde{B} & = r \left[ \frac{1}{8}\tilde{\rho} +  \lambda \left( I_{1}(a)+\frac{\Phi(a)}{12}\right)\frac{\left[r^3\mathcal{F}(r)\right]^\prime}{3r^2} \right]^\prime \\
\tilde{R} & = -\frac{\tilde{\rho}}{8}+\frac{\tilde{U}}{2} -  \lambda\left[I_{1}(a)+\frac{\Phi(a)}{12}\right] \frac{\left[r^3\mathcal{F}(r)\right]^\prime}{3r^2}  \\ & + \lambda\left[ I_{2}(a)+\frac{\Phi(a)}{6}\right]\mathcal{F}(r),  \\
\end{split}
\end{eqnarray}
where $\Phi$, $I_{1}$ and $I_{2}$ are three time dependent functions multiplying the anisotropic terms, and it is simple to 
see that when $\lambda=0$ one is recovering the isotropic limit of the quasi homogeneous solution. 

The effective curvature profile $\mathcal{K}(r)$ is now given by
\beq\label{Effective Curvature Profile2}
\mathcal{K}(r) \equiv K(r) - \frac{\lambda}{r^2_\mathrm{m}} \Phi(a)\mathcal{F}(r),
\eeq
where $\mathcal{F}(r)$ is defined as
\beq \label{F2}
\mathcal{F}(r) \equiv  r_\mathrm{m}\int_{\infty}^{r}  f(r^\prime) \mathrm{d}r^\prime \,.
\eeq
and in the \App{app:The pressure and energy density gradient profiles} one can find the details to compute explicitly the profile of $f(r)$, analyzing how this is varying with 
 $\lambda$.

The time dependent functions $\Phi$, $I_{1}$ and $I_{2}$, are solutions of the following system of equations:
\begin{align} 
\Phi^\prime(N) + 3\Phi(N) & = 3\frac{\rho^n_\mathrm{b}(N)}{a(N)r_\mathrm{m}} \nonumber \\
I^\prime_{1}(N) + 2I_{1}(N) & = -\frac{\Phi(N)}{6} - \frac{\rho^n_\mathrm{b}(N)}{2a(N)r_\mathrm{m}} \\ 
I^\prime_{2}(N) + 2I_{2}(N) & = - \frac{\Phi(N)}{3} \nonumber
\end{align}
where we have chosen $\Phi(0) = I_{1}(0) = I_{2}(0)=0$ as boundary conditions. This refers to the fact that at the initial time 
$N=0$, corresponding to an initial scale factor $a=a_\mathrm{i}$, when one assumes the perturbations have been 
generated - by inflation or any other physical mechanism in the very early Universe - it is reasonable to consider the radiation
medium to be still isotropic. The solution for $\Phi$, $I_{1}$ and $I_{2}$ obtained from the above mentioned system of differential 
equations is given 
by the following expressions:
\begin{align}\label{Phi+I1+I2 solutions}
\begin{split}
 \Phi(a) &= \Phi_0 \left(\frac{a}{a_\mathrm{i}}\right)^{-3}\left[ \left(\frac{a}{a_\mathrm{i}}\right)^{2(1-2n)} - 1 \right] \\
I_{1}(a) &=  - \frac{1}{6(1-4n)}\Phi_0 \left(\frac{a}{a_\mathrm{i}}\right)^{-3}  \left[ 1 - 4n  \right. \\
& \left. - 4(1-2n)\frac{a}{a_\mathrm{i}} + (3-4n)\!\left(\frac{a}{a_\mathrm{i}}\right)^{2(1-2n)}\right]  \\ 
I_{2}(a)  &= - \frac{1}{3(1-4n)} \Phi_0 \left(\frac{a}{a_\mathrm{i}}\right)^{-3} \left[1-4n  \right. \\  
&\left. - 2(1-2n)\frac{a}{a_\mathrm{i}}  + \left(\frac{a}{a_\mathrm{i}}\right)^{2(1-2n)}\right] . 
\end{split}
\end{align}
where
\beq
\Phi_0 = \frac{3}{2(1-2n)}\frac{\rho^n_\mathrm{b,i}}{a_\mathrm{i}r_\mathrm{m}} \,.
\eeq

When $n=1/2$ and $n=1/4$ the solution for $\Phi$, $I_1$ and $I_2$ requires some care (see \App{special_values} for 
more details).
The crucial difference between this case and to the previous one of \Sec{f=R}, is the presence of the function $\Phi(a)$ 
in \Eq{Effective Curvature Profile2}, which is also sourcing the functions $I_1(a)$ and $I_2(a)$. In general, these three 
functions are not dimensionless because of the time dependent coefficient  $\rho_\mathrm{b,i}^n(a)/(ar_\mathrm{m})$, 
changing the nature of the anisotropic parameter $\lambda$, which is also not dimensionless. This corresponds to a 
characteristic physical scale for the problem, as one can see in the definition of $\mathcal{F}$(r) in \Eq{F2} where 
the intrinsic scale $r_\mathrm{m}$ is now explicitly appearing. 

These functions modulate how the anisotropic behavior of the medium is varying during the expansion of the Universe, 
whereas in the previous case the anisotropy was independent of the Universe expansion. In the limit of $n=-1/4$ and $a \gg a_i$
the functions $\Phi$, $I_1$ and $I_2$ become time independent, and normalizing $\rho_\mathrm{b,i}^n(a)/(ar_\mathrm{m}) = 1$, 
we get $\Phi= \Phi_0=1$, $I_1 = - 1/3$ and $I_2 = - 1/6$, reproducing the solution of \Sec{f=R}.
\noindent

\section{Results}\label{results}

We can now compute explicitly the anisotropic initial conditions for different values of $\lambda$ in order to study the effect of 
the anisotropy on the shape of the energy density perturbation profiles, which will translate into a modified threshold for PBHs. 
With the fourth order Runge-Kutta numerical algorithm we compute the pressure and energy density gradient profiles  
(see \App{app:The pressure and energy density gradient profiles} for more details) enabling explicit computation of the quasi 
homogeneous solution derived in the previous section.  

\begin{figure*}[t!]
\centering
\vspace{-1cm}
\includegraphics[width=0.495\textwidth]{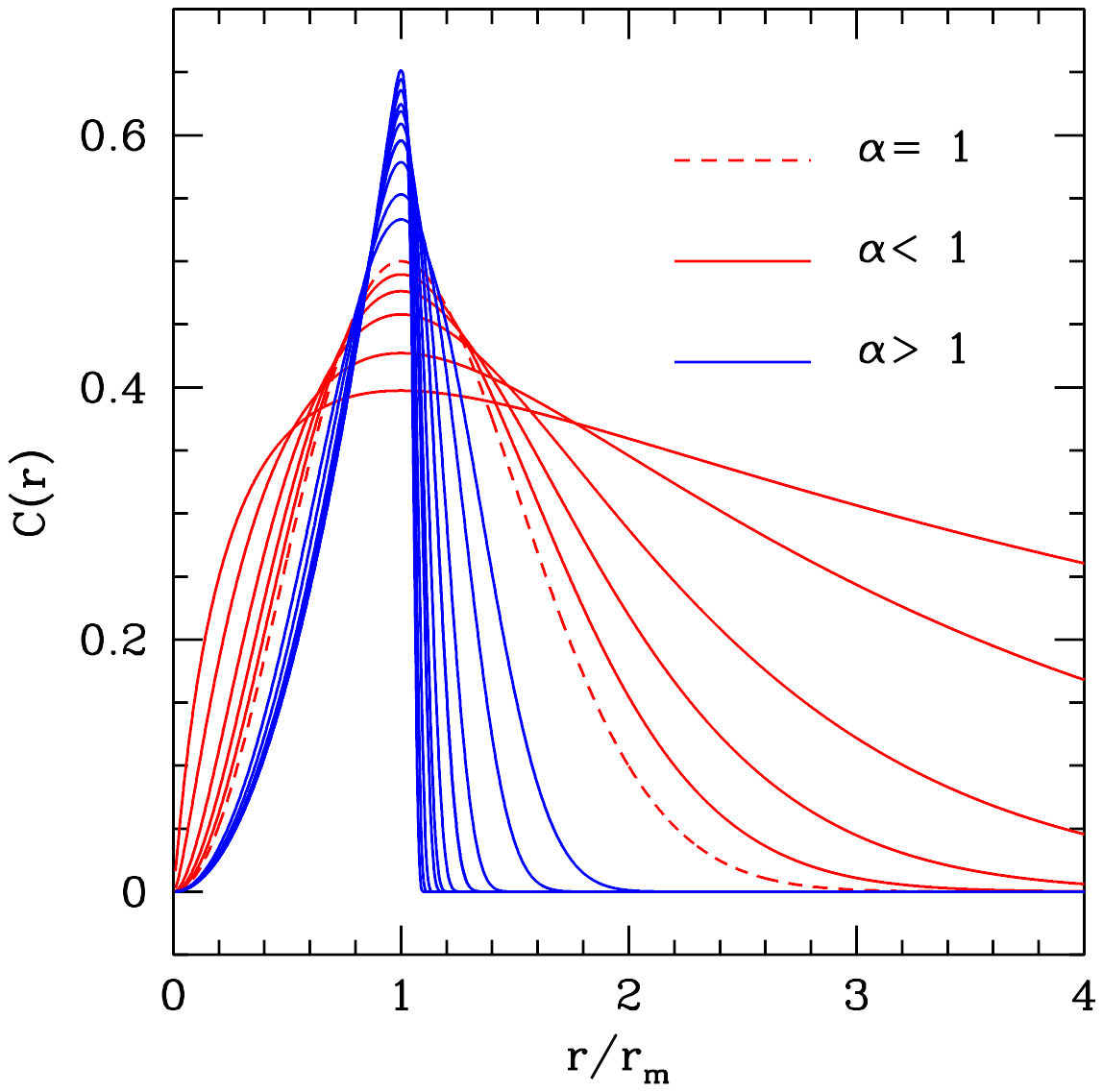}
\includegraphics[width=0.495\textwidth]{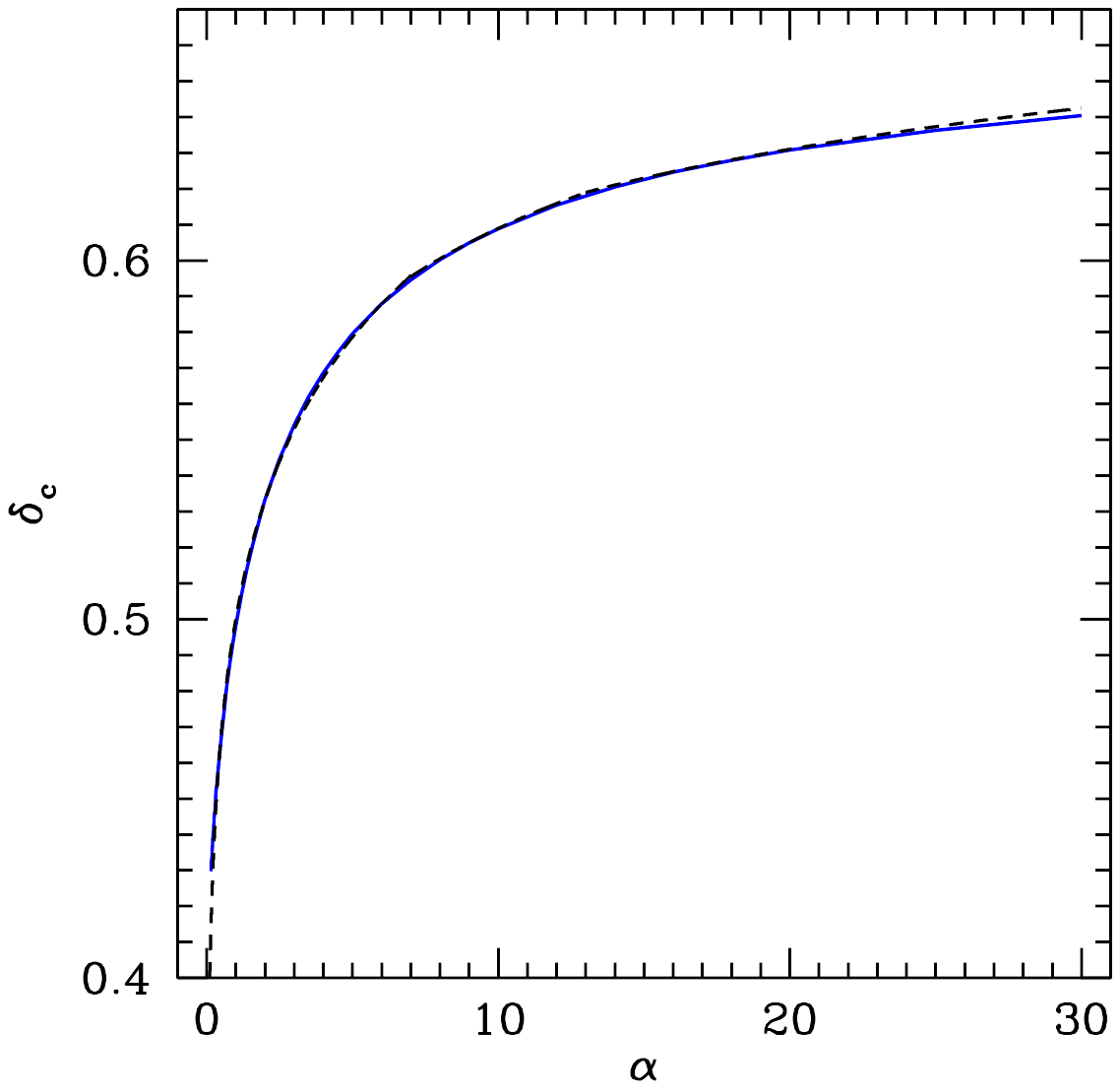}
\vspace{-2.5cm}
\caption{The left had plot shows the behavior of the compaction function varying the shape parameter $\alpha$ while the right panel
shows the numerical data for $\delta_c$, using a blue line, in terms of $\alpha$ while the analytic fit given by 
\eqref{delta_c_analytical_Musco} is plotted with a dashed line. In particular we are using here the curvature profile given by 
\eqref{K(r)} for $\lambda=0$.}
\label{fig:rho_alpha}
\end{figure*}

\subsection{The shape parameter}\label{shape}
As seen in \cite{Musco:2018rwt,Escriva:2019phb,Musco:2020jjb} the threshold for PBHs depends on the shape of the cosmological perturbation, characterized by the width of the peak of the compaction function $\mathcal{C}(r)$ defined 
in \Eq{C definition}, measured by a dimensionless parameter $\alpha$ defined as
\beq\label{alpha}
\alpha \equiv -\frac{r^2_\mathrm{m}\mathcal{C}^{\prime\prime}(r_\mathrm{m})}{4\mathcal{C}(r_\mathrm{m})} .
\eeq
The radius $r_\mathrm{m}$ is the characteristic comoving scale of the perturbation, identified where the compaction 
function has a peak, corresponding to the the location where the gravitational filed reaches its maximum. The apparent 
horizon of a black hole forms in this region during the collapse if the height of the peak, measuring the perturbation amplitude 
$\delta$, is larger than a threshold $\delta_\mathrm{c}$. 

For larger values of  $\alpha$ the peak of the compaction function becomes sharper while the peak of the energy density perturbation gets broader, whereas for smaller values of $\alpha$ we have the opposite behavior. The strict relation between the shape of the compaction function and the shape of the energy density perturbation is related to the Birkhoff theorem, where the collapse is mainly affected by the matter distribution inside the region forming the black hole, characterized just by one parameter, plus very small second order corrections induced by the shape of the perturbation outside this region~\cite{Musco:2018rwt}.

Looking at the quasi homogeneous solution derived in \Sec{initial} we have
\beq \label{Compaction}
\mathcal{C}(r) \simeq \frac{r^2}{r^2_\mathrm{m}}\tilde{M} +  O(\epsilon^2) = \frac{2}{3}\mathcal{K}(r)r^2
\eeq
which is a generalization of the expression for the isotropic solution, replacing $K(r)$ with $\mathcal{K}(r)$. The value of $r_\mathrm{m}$ is computed imposing  $\mathcal{C}^\prime(r_\mathrm{m})=0$, which gives 
\beq\label{r_m definition}
\mathcal{K}(r_\mathrm{m}) + \frac{r_\mathrm{m}}{2}\mathcal{K}^\prime(r_\mathrm{m}) = 0 .
\eeq
Using \Eq{Effective Curvature Profile2},  we can explicitly write \Eq{r_m definition} as
\beq\label{r_m2}
K(r_\mathrm{m}) + \frac{r_\mathrm{m}}{2} K^\prime(r_\mathrm{m}) =  \frac{ \lambda} {r^2_\mathrm{m}}  \Phi(a)
\left[ \mathcal{F}(r_\mathrm{m}) + \frac{r_\mathrm{m}}{2} \mathcal{F}^\prime(r_\mathrm{m})\right] . 
\eeq

To calculate the shape parameter $\alpha$ we insert \Eq{Effective Curvature Profile2} into 
\Eq{Compaction} and calculate the second derivative $\mathcal{C}^{\prime\prime}(r_\mathrm{m})$. 
The full expression for $\alpha$ in terms of $K(r)$, $\mathcal{F}(r)$, $\Phi(a)$, $\lambda$ and $j$  is very complicated, 
but we can understand the qualitative effect of the anisotropy by making a perturbative expansion for $\lambda \ll 1$
\beq\label{alpha perturbative}
\begin{split}
\alpha\simeq \alpha_0\left\{ 1 +  \left[p(r_\mathrm{m}) - q(r_\mathrm{m})\right] \Phi(a) \lambda  \right. \\ \left.  
  + p(r_\mathrm{m})q(r_\mathrm{m}) \Phi^2(a) \lambda^2  \right\},
\end{split}
\eeq
where  $\alpha_0$ is the shape parameter when $\lambda=0$, and $p(r)$ and $q(r)$ are two dimensionless functions 
defined as 
\begin{eqnarray}
p(r) & \equiv & \frac{\mathcal{F}(r)}{K(r)r^2} \label{p} \\ 
q(r) & \equiv & \frac{1}{r^2}\frac{\mathcal{F}^{\prime\prime}(r)r^2 + 4r\mathcal{F}^\prime(r)+2\mathcal{F}(r)}{K^{\prime\prime}(r)r^2+4rK^\prime(r) + 2K(r)}. \label{q} 
\end{eqnarray}

The shape parameter $\alpha_0$ of the isotropic solution is related to a family of curvature profiles $K(r)$ 
\beq \label{K(r)}
K(r) = \mathcal{A}\exp\left[ - \frac{1}{\alpha_0} \left(\frac{r}{r_\mathrm{m,0}}\right)^{2\alpha_0} \right],
\eeq
where $r_\mathrm{m,0}$ is the comoving scale of the perturbation, obtained from \Eq{r_m2}  when $\lambda=0$, 
and $\mathcal{A}$ is a parameter varying the perturbation amplitude $\delta$ as (see  \cite{Musco:2018rwt} for more details) 
\beq
\delta = \frac{2}{3} e^{-1/\alpha_0} \mathcal{A}\,r^2_\mathrm{m,0}  \,.
\eeq

The left plot of Figure~\ref{fig:rho_alpha} shows the compaction function profiles, obtained from \eqref{K(r)} when $\lambda=0$, 
for different value of $\alpha$. The peak of the compaction function becomes broader (red lines) for smaller values of $\alpha$, 
corresponding to a shape of the energy density profiles more and more peaked. Instead for larger values of $\alpha$ the 
compaction function is more peaked (blue lines) while the energy density profiles become broader. For $\alpha=1$ we have the
particular case of a Mexican hat shape for the energy density, obtained using a Gaussian profile for the curvature profile $K(r)$.

\subsection{The threshold for PBH formation}
\label{sec:The PBH formation threshold}

As the numerical simulations have shown, in a radiation dominated Universe there is a simple analytic relation for the 
threshold of PBH formation as a function of the shape parameter, $\alpha$, corresponding to the numerical fit given by 
Eq. (44) of~\cite{Musco:2020jjb}:
\beq\label{delta_c_analytical_Musco}
\delta_\mathrm{c}= 
\begin{cases}
\alpha^{0.047}-0.50 \quad\quad\quad 0.1 \lesssim\alpha\lesssim 7 \\
\alpha^{0.035}-0.475 \quad\quad\quad 7 \lesssim\alpha\lesssim 13 \\
\alpha^{0.026}-0.45 \quad\quad\quad 13 \lesssim\alpha\lesssim 30.
\end{cases}
\eeq
This is represented in the right plot of Figure~\ref{fig:rho_alpha}, where the numerical data is plotted with a blue line, while the fit 
given by \eqref{delta_c_analytical_Musco} is plotted with a dashed line.

\begin{figure*}[ht!]
\centering
\includegraphics[width=0.497\textwidth]{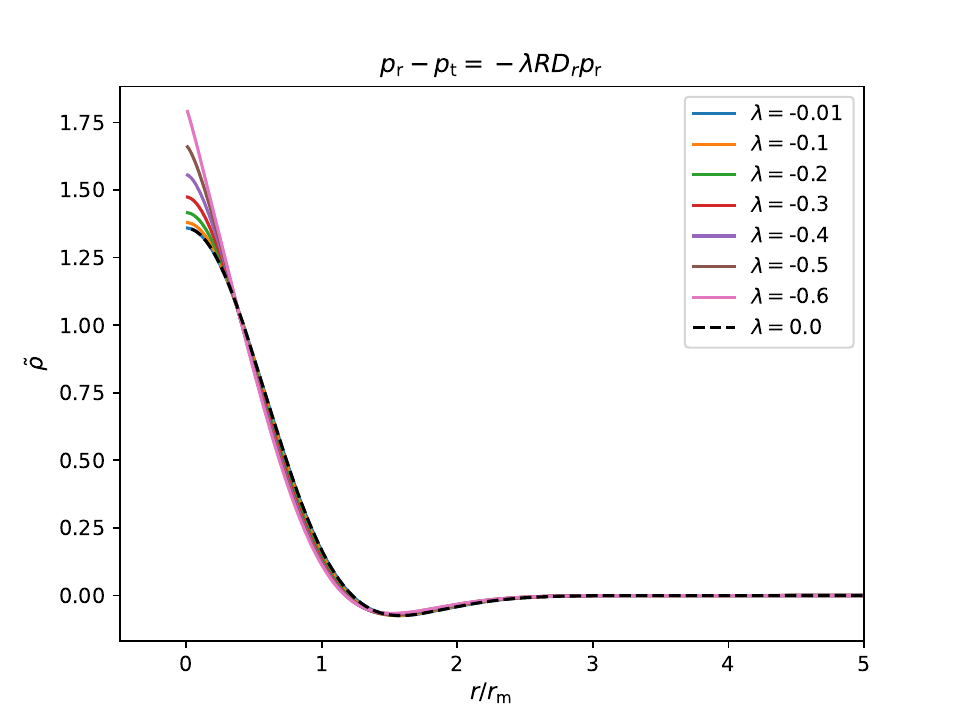}
\includegraphics[width=0.497\textwidth]{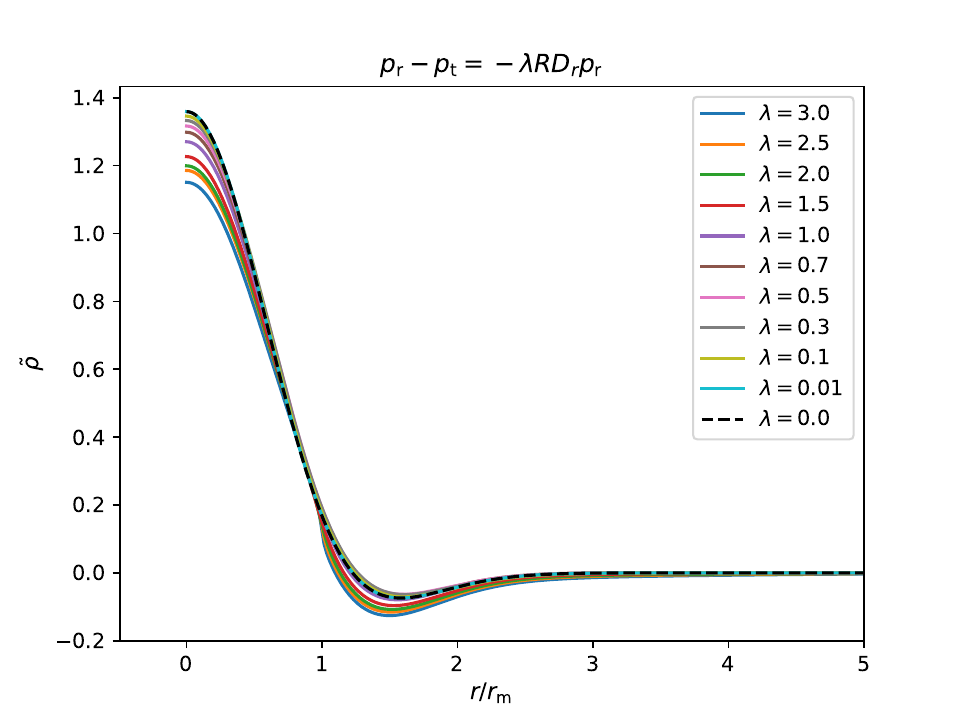}
\includegraphics[width=0.497\textwidth]{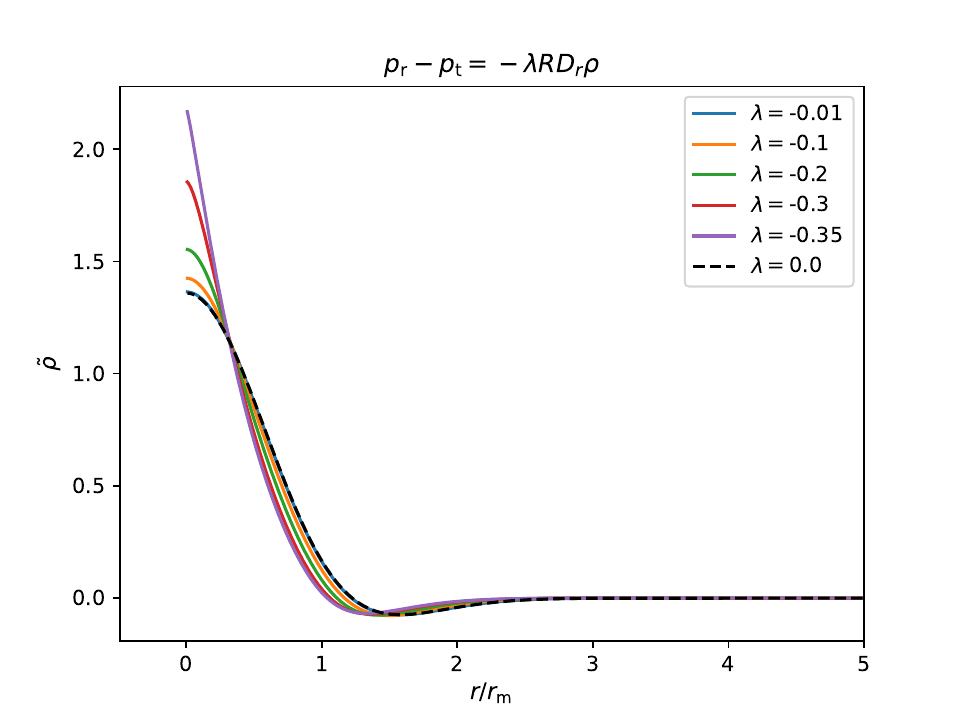}
\includegraphics[width=0.497\textwidth]{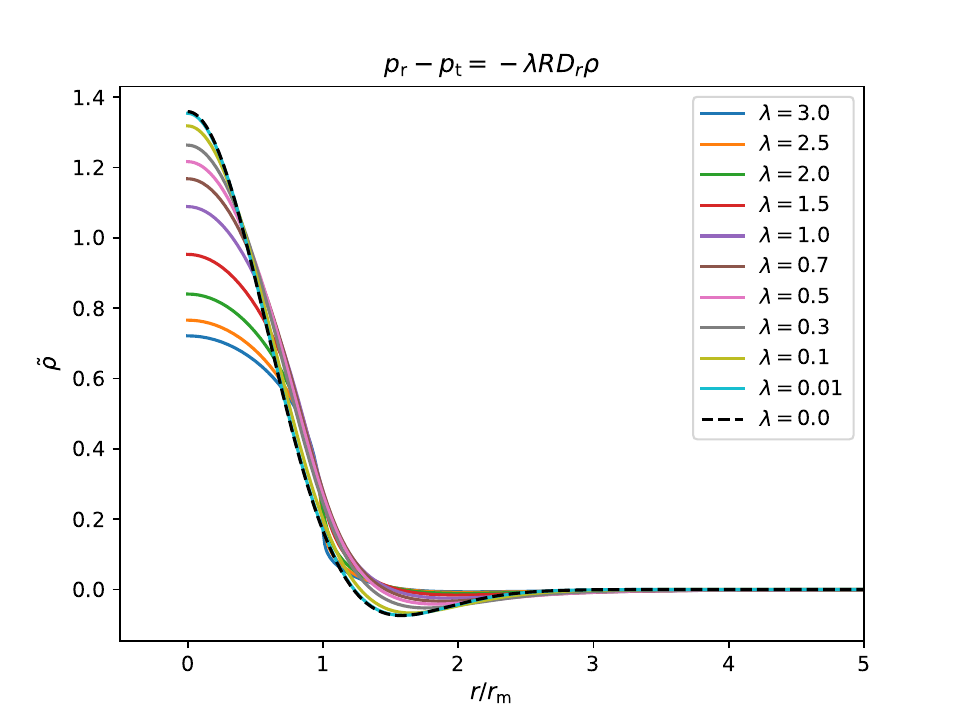}
\caption{In this figure, we show the behavior of $\tilde{\rho}$ plotted against $r/r_\mathrm{m}$ for the special case $n=0$ from 
the family of models with $g(r,t)=R(r,t)$.  In the top panels, we consider the case where the equation of state of the anisotropic 
fluid is given in terms of pressure gradients, following \Eq{eq:p_r+p_t - f= R-D_rp_r}, whereas in the bottom panels we account 
for the case where the equation of state is given in terms of energy density gradients, following \Eq{eq:p_r+p_t - f= R-D_rrho}. 
The left panels consider negative values of the anisotropy parameter $\lambda$ while the right ones account for positive values.}
\label{fig:rho_tilde_f=R}
\end{figure*}

Inserting \eqref{K(r)} into \Eq{Effective Curvature Profile2}, and solving numerically for the function $\mathcal{F}(r)$ 
(see \App{app:The pressure and energy density gradient profiles}) to compute the profiles of the pressure or energy density 
gradients, we can study how the Mexican hat profile of the energy density, taken as a typical perturbation, is modified by the 
anisotropy, varying $\lambda$. To do this we consider a constant value of the perturbation amplitude  $\delta=0.5$, taking into 
account that $\delta_\mathrm{c} \simeq 0.5$ is the threshold for the Mexican hat shape ($\alpha_0=1$ and 
$\lambda=0$). The relation between $\delta_\mathrm{c}$ and $\alpha$ given in \eqref{delta_c_analytical_Musco} then allows the 
corresponding value of the threshold to be computed in terms of $\lambda$. 

Here we are assuming that the effect of the anisotropy could be computed with the non linear
modification of the shape, without modifying the relation between the shape and the threshold. This is a reasonable approximation without performing full non linear simulations of the anisotropic collapse. 

After normalizing  $r^2_\mathrm{m,0}=1$ and inserting this into \Eq{r_m2} we find that $r_\mathrm{m} \simeq r_\mathrm{m,0}$
which means that there is no a significant change in the characteristic scale because of the anisotropy. The main effect on the shape is given by the competition of the two functions $p(r)$ and $q(r)$ defined in \Eqs{p} and \eqref{q}. In general we have observed that
$p(r_\mathrm{m})>q(r_\mathrm{m})$ and therefore from \Eq{alpha perturbative} one can easily infer that, considering terms 
up to order $O(\lambda)$ in \Eq{alpha perturbative}, for positive values of $\lambda$ the value of the shape parameter 
$\alpha$ increases, making the shape of the compaction function sharper while the shape of the energy density perturbation profile becomes broader. On the other hand, negative values of $\lambda$ give a smaller value of $\alpha$, broadening the shape of the compaction function while the energy density perturbation profile gets steeper. 

This behavior is shown explicitly in Figure~\ref{fig:rho_tilde_f=R}, where we plot $\tilde{\rho}$ for different values of  
$\lambda$ when $g(r,t)=R(r,t)$: the upper plots correspond to the EoS expressed in terms of pressure gradients ($j=0$) 
while the bottom ones refer to the EoS expressed in terms of energy density gradients ($j=1$). The left plots of this figure 
are characterized by negative values of $\lambda$ while the right plots are characterized by positive values of 
$\lambda$.  

Starting from $\lambda = 0$ when the fluid is isotropic, we observe for $\lambda<0$ an increase of the amplitude of the 
peak of the energy density perturbation and the central profile sharpens more and more, consistently with the increase of the pressure gradients in the center observed in the left panels of Figure~\ref{fig:rho_tilde_f=R}. This translates into a broadening of the peak of the compaction function, decreasing the value of $\delta_\mathrm{c}$ and enhancing in this way the formation of PBHs. 

This could be explained with simple physical arguments by the following reasoning: given the fact that the pressure/energy density gradient profile is mainly negative (see \App{app:The pressure and energy density gradient profiles}), from $p_\mathrm{r}-p_\mathrm{t}=-\lambda RD_\mathrm{r}\left( p_\mathrm{r}\rm{\;or\;} \rho \right)$, one has that $p_\mr<p_\mt$ and the radial pressure is reduced with respect to the tangential one. Because of this, one would expect it to 
be easier for a cosmological perturbation to collapse along the radial direction with respect to the isotropic case and consequently the peak of the energy density perturbation to be larger compared to the isotropic case with $\lambda=0$. 

On the other hand, when $\lambda>0$ we have $p_\mr>p_\mt$, giving a larger value of the radial component of the 
pressure compared to the isotropic case. In this case the pressure gradients are increased around $r_\mathrm{m}$ as 
shown in the right panels of Figure~\ref{fig:rho_tilde_f=R}. This translates into a reduced amplitude of the peak of 
the energy density perturbations with respect to the isotropic case, which makes the collapse of cosmological perturbations 
into PBHs more difficult, increasing consequently the value of $\delta_\mathrm{c}$.

\begin{figure*}[ht!]
\centering
\includegraphics[width=0.497\textwidth]{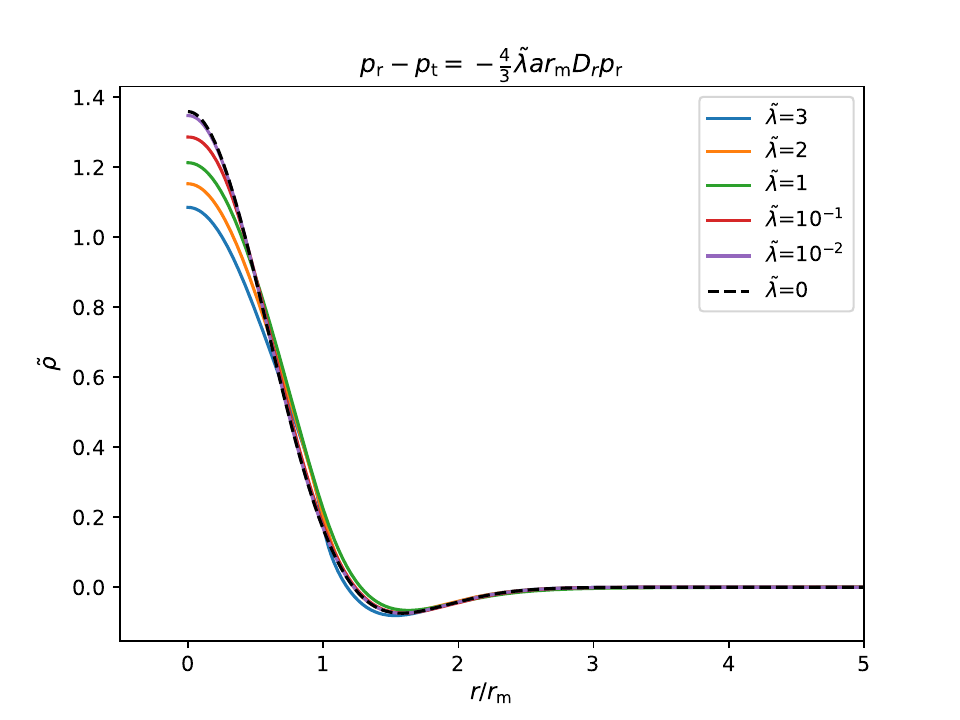}
\includegraphics[width=0.497\textwidth]{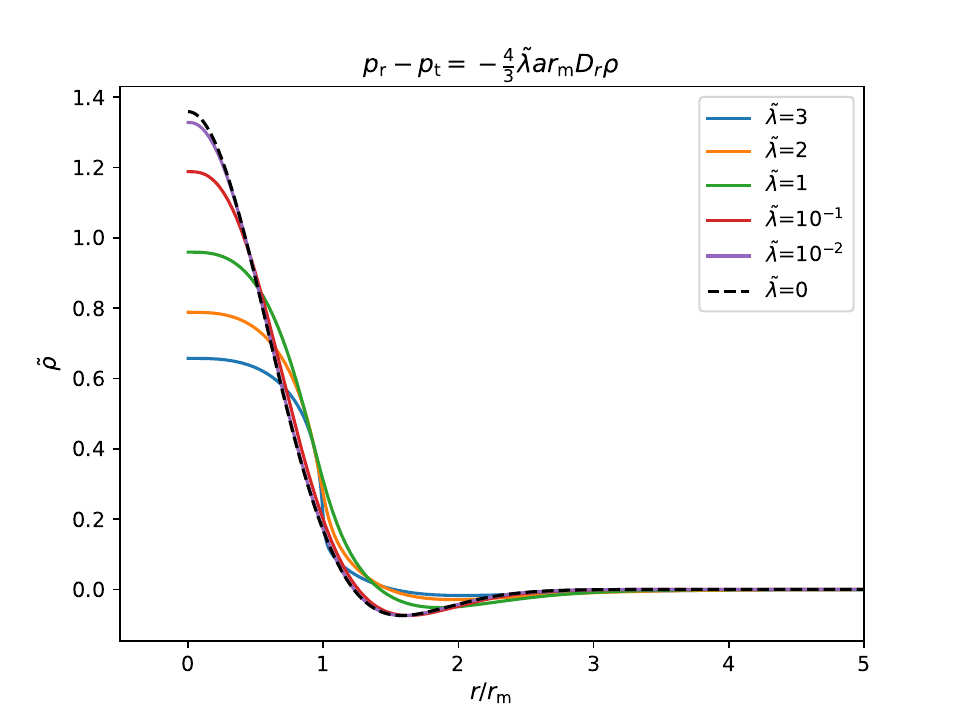}
\caption{In this figure, we show the behavior of $\tilde{\rho}$ against $r/r_\mathrm{m}$ when $g(r,t)=\rho^n(r,t)$. In the left panel, we consider the case where the equation of state of the anisotropic fluid  is given in terms of pressure gradients, following \Eq{eq:p_r+p_t - f= rho^n-D_rp_r}, whereas in the right panel we account for the case where the equation of state is given in terms of energy density gradients, following \Eq{eq:p_r+p_t - f= rho^n-D_rrho}.}
\label{fig:rho_tilde_pr_rho_f=rho^n}
\end{figure*}

In Figure~\ref{fig:rho_tilde_pr_rho_f=rho^n}, we analyze the effect of the anisotropy on the profile of the energy density perturbation when the equation of state is characterized by $g(r,t)=\rho^n(r,t)$, rescaling the anisotropic parameter, measured at horizon crossing $a_\mathrm{HC}$, in a dimensionless form 
\beq\label{tilde_lambda}
\tilde{\lambda}\equiv \lambda\Phi(a_\mathrm{HC}),
\eeq
which allows he EoS to be rewritten as
\beq\label{p_r-p_t-tilde_lambda-f=rho^n}
p_\mathrm{r} = \frac{1}{3}\left[\rho-2\tilde{\lambda}r_\mathrm{m}\chi_n(a)\left(\frac{\rho}{\rho_\mathrm{b,i}}\right)^n
D_\mathrm{r} \left\{ 
\begin{split} 
& p_\mathrm{r} &(j=0) \\ 
& \rho  &(j=1)
\end{split}
\right\} \right],
\eeq
where
\beq
\chi_n(a)\equiv \frac{2a_\mathrm{i}(1-2n)}{3}\frac{\displaystyle{\left(\frac{a}{a_\mathrm{i}}\right)^3}}{\left[\displaystyle{\left(\frac{a}{a_\mathrm{i}}\right)^{2(1-2n)}-1}\right]} \,.
\eeq
In this case, we consider only positive values of $\tilde{\lambda}$ because of the structure of the equations for the 
pressure or energy density gradients (see \Eq{p_r ODE compact form - f=rho^n} and \eqref{rho ODE compact form - f=rho^n} 
in \App{A2}). 

As we have discussed in Section~\ref{f=rho^n}, this EoS introduces a characteristic scale into the problem, which requires 
specification of an additional parameter \mbox{$\mu \equiv \left(\frac{\rho_\mathrm{b,HC}}{\rho_\mathrm{b,i}}\right)^{1/4}$} \,, defined as 
the ratio between the energy scales at horizon crossing  ($\epsilon_\mathrm{HC}=1$) and at the initial time $t_i$, when the 
perturbations are generated. This time depends on the particular cosmological model of the early Universe being considered 
(e.g. inflation).

From the EoS seen in \Eq{p_r-p_t-tilde_lambda-f=rho^n} one can identify three main contributions: the dimensionless parameter $\tilde{\lambda}$ accounting for the anisotropy of the medium, the term $\left(\frac{\rho}{\rho_\mathrm{b,i}}\right)^n$ measuring the effect of cosmic expansion, and finally $D_\mathrm{r}\left(p_\mathrm{r}\rm \ \textrm{or} \ \rho\right)$ which accounts for the effect of the pressure or energy density gradients. As it seems reasonable, we assume that for $t\to\infty$ the contribution of the pressure and energy density gradients disappears. This constrains the value of the exponent of $\rho$ just to non negative values ($n\geq 0$) and it is interesting to note that this is discarding the solution analyzed in \Sec{f=R}. 

In Figure~\ref{fig:rho_tilde_pr_rho_f=rho^n} we analyze the simplest model with $n=0$ and $\mu=10^{-10}$. The qualitative behavior is similar to the case where $g(r,t)=R(r,t)$ with a positive value of $\tilde{\lambda}$ enhancing the radial pressure compared to the tangential one and reducing the height of the peak of $\tilde{\rho}$ with respect the isotropic case, making in this way more difficult for cosmological perturbations to collapse into PBHs. This is confirmed by the behavior of the pressure
gradients seen in \Fig{fig:dpr_dr+drho_dr_f=rho^n}, similar to the one seen in the right plots of \Fig{fig:dpr_dr+drho_dr:f=R}.


\begin{figure*}[ht!]
\centering
\includegraphics[width=0.497\textwidth]{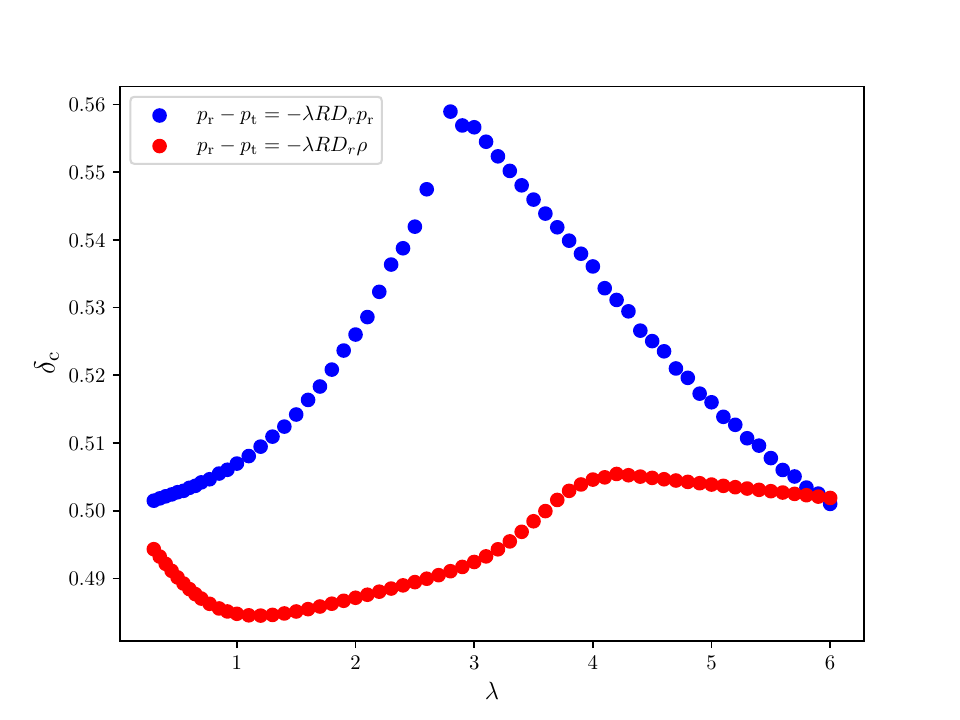}
\includegraphics[width=0.497\textwidth]{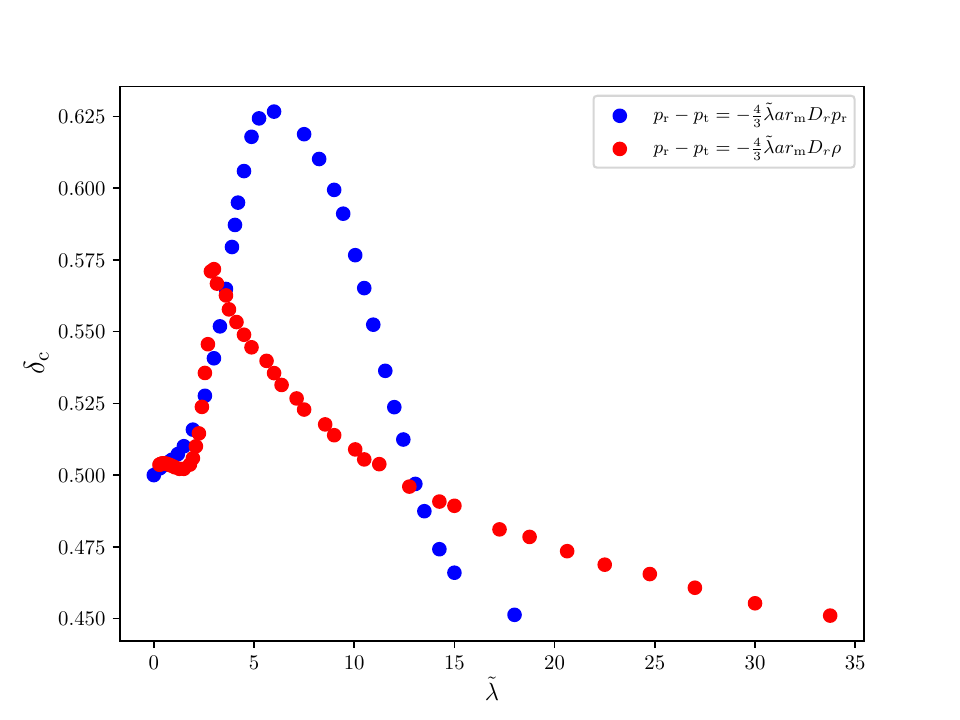}
\caption{This figure shows the threshold of PBHs $\delta_\mathrm{c}$ as a function of the amplitude of the anisotropy in linear 
scale.  In the left panel, we see the case where $g(r,t)=R(r,t)$ while in the right panel we consider a more general model with 
$g(r,t)=\rho^n(r,t)$. For both cases the blue dots indicate the values of the threshold when the anisotropic term of the equation 
of state is modeled in terms of pressure gradients while the red dots correspond to values of the threshold when the anisotropic
term is written in terms of gradients of the energy density.}
\label{fig:delta_c_lambda}
\end{figure*}

\begin{figure*}[ht!]
\centering
\includegraphics[width=0.497\textwidth]{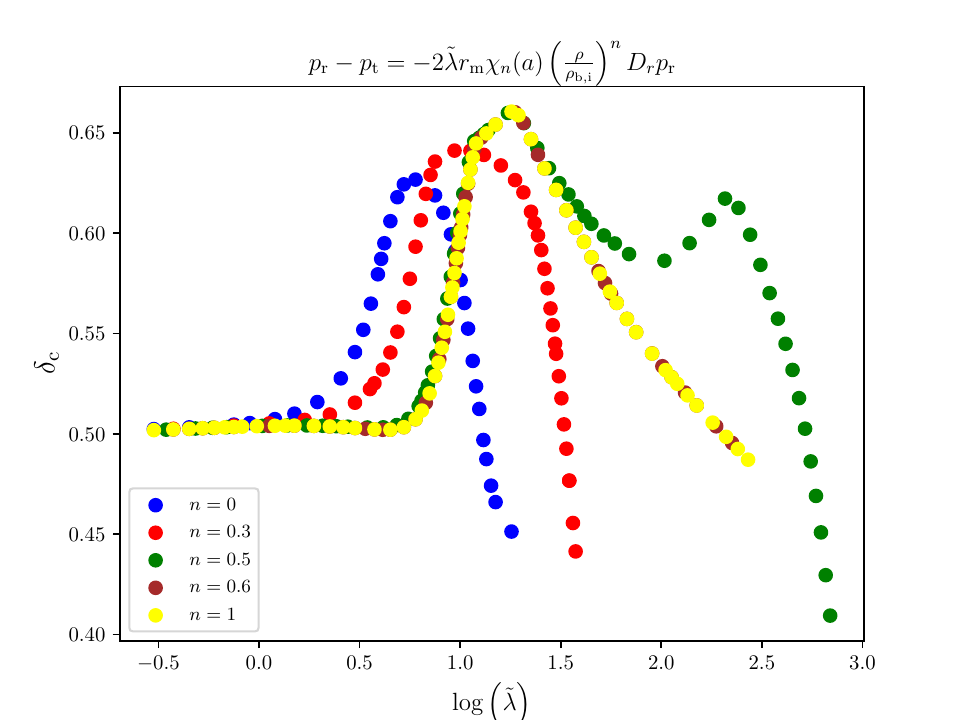}
\includegraphics[width=0.497\textwidth]{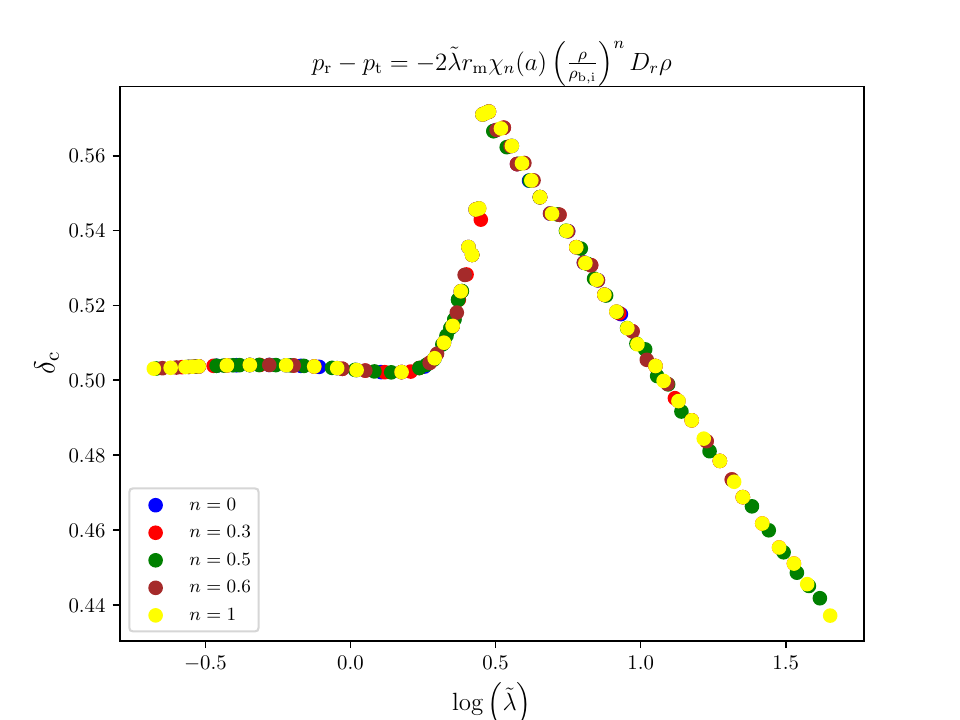}
\caption{This figure shows the threshold of PBHs $\delta_\mathrm{c}$ as a function of the amplitude of the anisotropy 
using when the equation of state is characterized by $g(r,t)=\rho^n(r,t)$ for $n=0, 0.3, 0.5, 0.6, 1$. In the left panel we see 
the case of the equation of state having an implicit form when this expressed in terms of pressure gradients, while in the 
right panel this has an explicit form as it is expressed in terms of density gradients.}
\label{fig:delta_c_lambda_n}
\end{figure*}

The effect of the anisotropy on the shape of the energy density perturbation can be used to estimate the corresponding effect on
the threshold $\delta_\mathrm{c}$ for PBH formation. To do so, we make the assumption that $\delta_\mathrm{c}$ has the same 
dependence on the shape of the initial energy density perturbation profile seen in the isotropic case, as given by 
\eqref{delta_c_analytical_Musco}. This enables us to study how $\delta_\mathrm{c}$ is 
varying with respect to the amplitude of the anisotropy, as shown explicitly in \Fig{fig:delta_c_lambda}, both for the model plotted in 
\Fig{fig:rho_tilde_f=R}  when $g(r,t) = R(r,t)$ (left panel) and for the model of \Fig{fig:rho_tilde_pr_rho_f=rho^n}, when 
$g(r,t)=\rho^n(r,t)$, using in particular $n=0$ and $\mu=10^{-10}$ (right panel). Finally in \Fig{fig:delta_c_lambda_n} we study the 
behavior of $\delta_c$ when $g(r,t)=\rho^n(r,t)$ for different values of $n$, considering in the left panel the EoS written in terms of 
pressure gradients while in the right one the EoS written in terms of density gradients is used.

In general we observe an initial increase of $\delta_\mathrm{c}$ with respect to $\lambda$ or $\tilde{\lambda}$, which is somehow 
expected, as already explained, because the shape parameter $\alpha$ becomes larger for an increasing amplitude of the anisotropy,
enhancing the radial pressure with respect to the tangential one. However from these figure we can see a critical value of $\lambda$ 
and $\tilde{\lambda}$, followed by a decreasing behavior of $\delta_\mathrm{c}$, when the modification of the shape parameter due to 
the anisotropy is non linear. This effect is due to the term $O(\lambda^2)$ in \Eq{alpha perturbative}, becoming important when 
$\lambda\sim1$.  Obviously this regime is challenging the validity of our approximation of computing the threshold 
using the isotropic relation between $\delta_c$ and $\alpha$, and this result should therefore be considered with care.

\Fig{fig:delta_c_lambda_n} shows that, while the model in terms of pressure gradients has a different behavior for different values of 
$n$, the model of the EoS written in terms of density gradients shows a universal behavior, independent of the particular value of $n$. 
This difference can be explained by the implicit solution of the equation of state, when this is expressed in terms of the pressure 
gradients with respect to the explicit form which has when written in terms of the density gradients.

Although these results are genuinely interesting and find a clear physical explanation, we stress again that one cannot fully trust the 
perturbative approach in the regime where $\delta_\mathrm{c}$ is decreasing and full numerical simulations solving the non-linear 
hydrodynamic equations are necessary to confirm to which extent \Eq{delta_c_analytical_Musco} holds for a non linear amplitude 
of the anisotropy. 

Despite this, the results obtained here give a reasonable estimation of the effect of the anisotropy on the threshold of PBH formation 
when the anisotropy is not too large, with a change of the threshold up to $25\%$. This would mean, potentially, a relevant change 
of the abundance for PBHs if the early Universe was significantly non isotropic. 
 

\section{Conclusions}\label{conclusions}
In this work, we have studied the formation of PBHs within a radiation fluid described by an anisotropic pressure. By making use of 
a covariant formulation of the equation of state and performing a gradient expansion approximation on superhorizon scales we have 
computed the anisotropic quasi homogeneous solution describing the initial conditions that one would need to use in the future for 
numerical simulations. Using this solution we have investigated the effect of the anisotropy on the shape of the energy density 
perturbation profile, estimating the corresponding value of the threshold for PBHs, assuming that $\delta_\mathrm{c}$ has the same behavior with the shape of the energy density profile as when the fluid is isotropic.

Although the estimation of the threshold for PBH computed here is consistent only for small values of the anisotropy parameter 
($\lambda\ll 1$), the qualitative behavior found for $\delta_\mathrm{c}$ looks to be consistent, and gives a reasonable solution 
to a problem that has never been studied before. To obtain a more quantitative and precise answer to such a problem, when the 
amplitude of the anisotropy is not small, it would be necessary to perform full numerical simulations, generalizing for example the 
code used in previous works of this type, as in~\cite{Musco:2004ak,Polnarev:2006aa,Musco:2008hv,Musco:2012au,Musco:2018rwt,Musco:2020jjb}. 

Before concluding we should comment here on the model with $g(r,r) = R(r,t)$ where the behavior of $\alpha$ and $\delta_\mathrm{c}$ 
is significantly different when $p_\mr-p_\mt$ is proportional to the pressure gradients from the  case when it is proportional to the 
energy density gradients. In the first case $\delta_\mathrm{c}$ is initially increasing with $\lambda$ up to a critical point and then
decreases, while in the second case $\delta_\mathrm{c}$ is first decreasing and then increasing. It is difficult to understand
the physical motivation of this discordant behavior. This model however is not based on solid physical grounds, because the 
EoS with $g(r,t)=R(r,t)$ is not expressed in terms of local quantities, as one would normally expect. This is a special case of the 
model described in \Sec{f=rho^n}, with $n=-1/4$, where the pressure or energy density gradients do not vanish for an infinite 
expansion, as one would expect.

Analyzing this first model has been useful to simplify the problem, understanding how to write the anisotropic  quasi homogeneous 
solution in a clear and self consistent form. However, only the model elaborated later in \Sec{f=rho^n}, where the EoS is written only 
in terms of local quantities, and the anisotropy is varying also with the expansion of the Universe, looks to be physically 
plausible, and therefore should be seriously taken into account for further studies on the subject, with particular attention to the version 
where the EoS is written in terms of density gradients, characterized by an explicit solution.

\section*{Acknowledgments}
We would like to thank Antonio W. Riotto, Paolo Pani, David Langlois, Vincent Vennin, Valerio De Luca, Gabriele Franciolini and John C. Miller,  for useful discussions and comments.

I.~M. acknowledges financial support provided under the European Union's H2020 ERC, Starting Grant agreement no.~DarkGRA--757480 
and under the MIUR PRIN programme, and support from the Amaldi Research Center funded by the MIUR program ``Dipartimento di 
Eccellenza" (CUP:~B81I18001170001).

T.~P. acknowledges financial support from the Fondation CFM pour la Recherche in France, the Alexander S. Onassis Foundation - Scholarship ID: FZO 059-1/2018-2019, the Foundation for Education and European Culture in Greece and the A.G. Leventis Foundation. 

\begin{appendix}
\section{EoS with {$g(r,t)=\rho^n(r,t)$} and $n=1/2,1/4$}\label{special_values}
Here we discuss the particular cases when $n=1/2$ and $n=1/4$. Looking at  \Eq{Phi+I1+I2 solutions}, one can see that the 
functions $\Phi$ , $I_1$ and $I_2$ diverge due to the prefactor $1/(1-2n)$ in  $\Phi$ and  $1/\left[(1-2n)(1-4n)\right]$ in $I_1$ and $I_2$. However, computing carefully these limits for $n\to 1/2$ one gets

\begin{align}
\Phi(a) & = \frac{3\sqrt{\rho_\mathrm{b,i}}}{a_\mathrm{i}r_\mathrm{m}}\left(\frac{a}{a_\mathrm{i}}\right)^{-3}\ln\left(\frac{a}{a_\mathrm{i}}\right) \\ 
I_{1}(a) & = \frac{\sqrt{\rho_\mathrm{b,i}}}{2a_\mathrm{i}r_\mathrm{m}}\left(\frac{a}{a_\mathrm{i}}\right)^{-3}\left[2-2\frac{a}{a_\mathrm{i}}+\ln\left(\frac{a}{a_\mathrm{i}}\right)\right] \\
I_{2}(a) & =  \frac{\sqrt{\rho_\mathrm{b,i}}}{a_\mathrm{i}r_\mathrm{m}}\left(\frac{a}{a_\mathrm{i}}\right)^{-3}\left[\ln\left(\frac{a}{a_\mathrm{i}}\right)-\frac{a}{a_\mathrm{i}}+1 \right]\ \,,
\end{align}
while for $n\to 1/4$ one has
\begin{align}
I_{1}(a) & = \frac{\rho^{1/4}_\mathrm{b,i}}{2a_\mathrm{i}r_\mathrm{m}}\left(\frac{a}{a_\mathrm{i}}\right)^{-3}
\left\{ \frac{a}{a_\mathrm{i}}\left[ 1 - 2\ln\left(\frac{a}{a_\mathrm{i}}\right) \right] - 1 \right\} \\
I_{2}(a) & = \frac{\rho^{1/4}_\mathrm{b,i}}{a_\mathrm{i}r_\mathrm{m}}\left(\frac{a}{a_\mathrm{i}}\right)^{-3} 
\left\{ \frac{a}{a_\mathrm{i}} \left[1-\ln\left(\frac{a}{a_\mathrm{i}}\right)\right] -1 \right\} \,.
\end{align}

\section{Density  and pressure gradients} \label{app:The pressure and energy density gradient profiles}
In this appendix, we give some additional details concerning the pressure and energy density gradient profiles for both EoSs, $g(r,t)=R(r,t)$ and $g(r,t)=\rho^n(r,t)$.

\subsection{Equation of state with $g(r,t)=R(r,t)$} \label{A1}
In the case where the equation of state is given in terms of pressure gradients, following \Eq{eq:p_r+p_t - f= R-D_rp_r}, in order to 
get $\frac{\partial\tilde{p}_\mathrm{r}}{\partial r}$, one should combine \Eq{p_tilde-rho_tilde} and the equation for $\tilde{\rho}$ from 
\Eq{Perturbations} to find the behavior of $f(r)$ defined in \Eq{f_j:f=R} as solution of  the following differential equation:
\beq\label{p_r ODE compact form}
\begin{aligned}
&\frac{4\lambda}{3}r\sqrt{1-K(r)r^2}f^\prime(r) + \left[\frac{7\lambda}{3}\sqrt{1-K(r)r^2} + \frac{3}{2} \right]f(r)
\\ & - \left\{\frac{\left[r^3K(r)\right]^\prime}{3r^2}\right\}^\prime r^2_\mathrm{m} \sqrt{1-K(r)r^2} = 0,
\end{aligned}
\eeq
with the boundary condition $f(0)=0$, as imposed by \Eq{boundary condition}. For $\lambda=0$ one recovers the isotropic quasi-homogeneous limit,
\beq\label{f(r) - lambda =0}
f_{\lambda=0}(r)= \frac{2}{3}\left\{\frac{\left[r^3K(r)\right]^\prime}{3r^2}\right\}^\prime r^2_\mathrm{m} \sqrt{1-K(r)r^2}.
\eeq
Solving \Eq{p_r ODE compact form} for $f(r)$, which allows to compute explicitly $\mathcal{F}(r)$ in \Eq{F:f=R}, one obtains the 
explicit for of the quasi homogeneous solution given in \eqref{Perturbations} written in terms of a given curvature profile $K(r)$.

\begin{figure*}[ht!]
\centering
\includegraphics[width=0.497\textwidth]{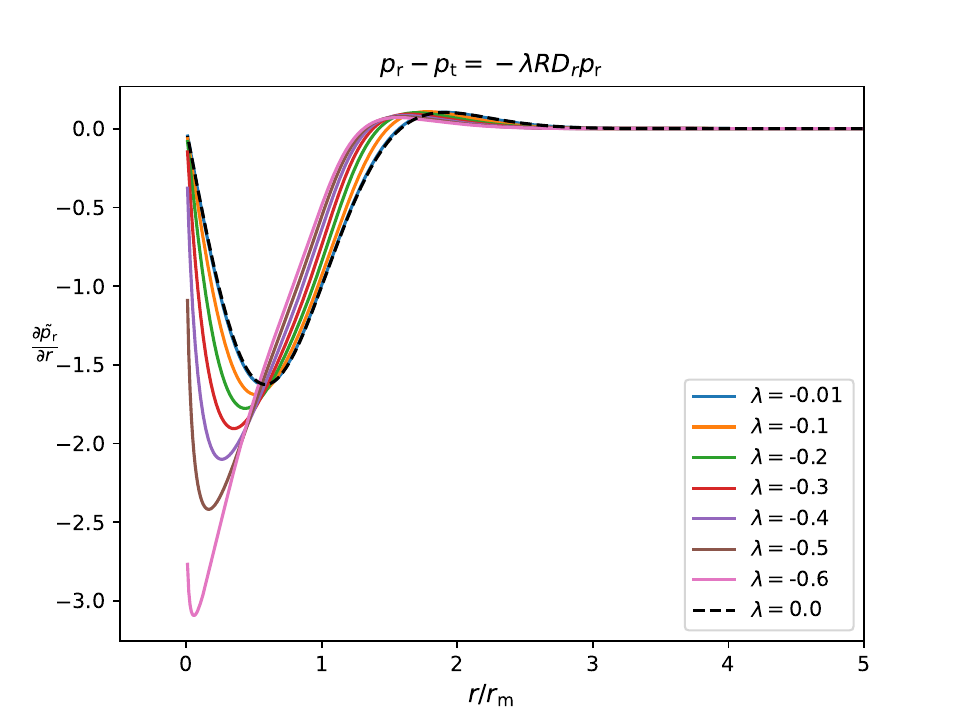}
\includegraphics[width=0.497\textwidth]{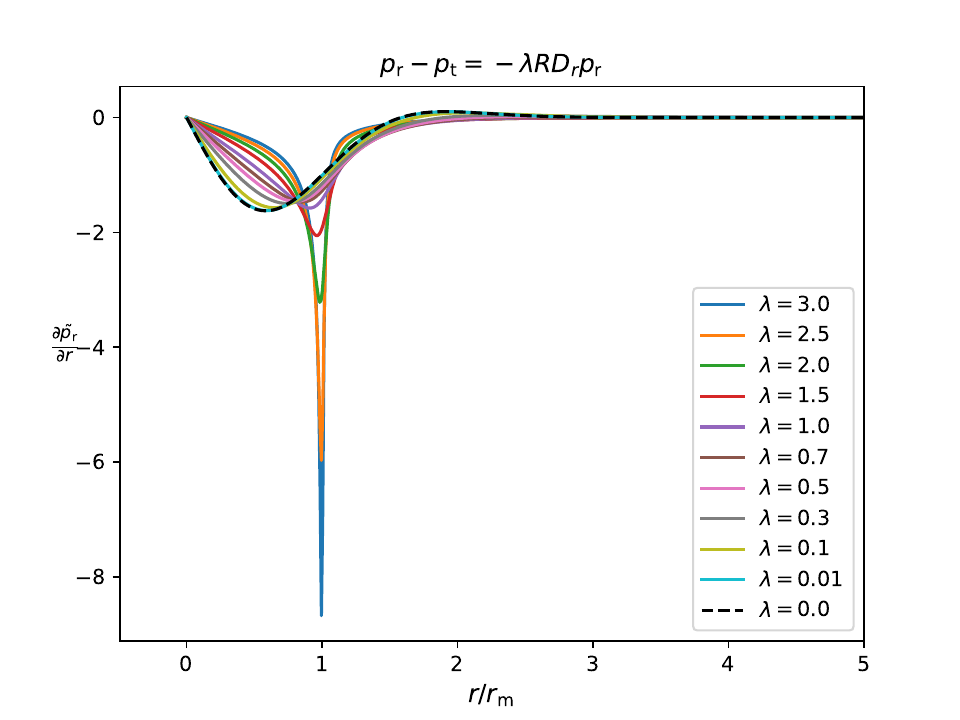}
\includegraphics[width=0.497\textwidth]{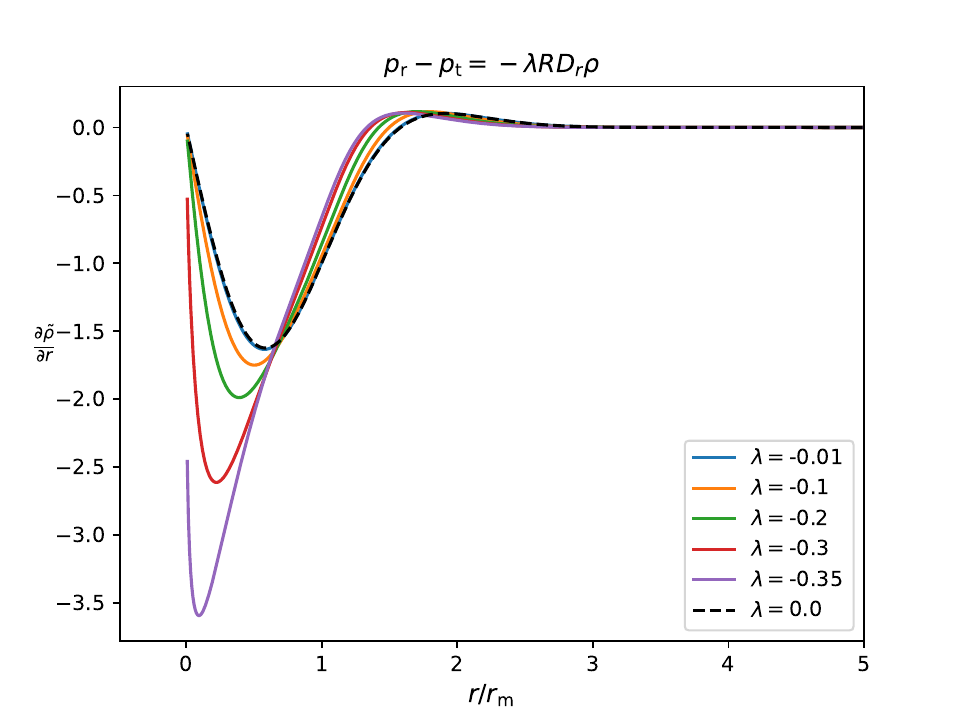}
\includegraphics[width=0.497\textwidth]{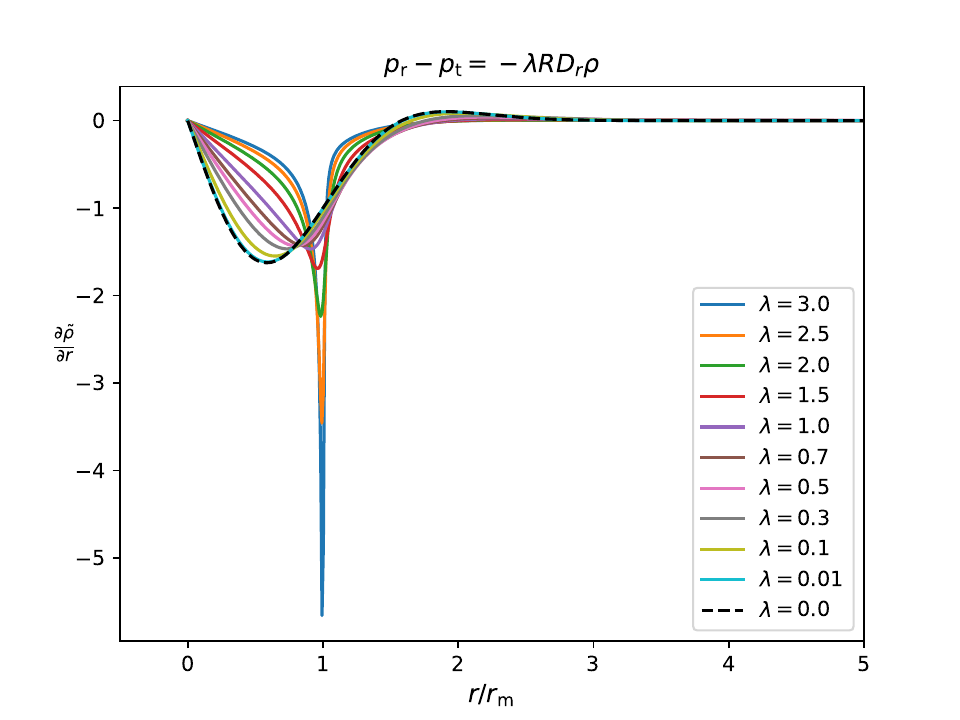}
\caption{In this figure, we show the behavior of $\frac{\partial\tilde{p}_\mr}{\partial r}$  and  $\frac{\partial\tilde{\rho}}{\partial r}$  
plotted against $r/r_\mathrm{m}$. The top panels concern the case where the equation of state is given in terms of pressure gradients, 
following \Eq{eq:p_r+p_t - f= R-D_rp_r}, whereas the bottom panels are for the case where the equation of state is given in terms of 
energy density gradients, following \Eq{eq:p_r+p_t - f= R-D_rrho}. The left figures show the gradient profiles when $\lambda<0$ whereas the right ones consider values of $\lambda>0$.}
\label{fig:dpr_dr+drho_dr:f=R}
\end{figure*}

In the case where the equation of state is given in terms of energy density gradients, following \Eq{eq:p_r+p_t - f= R-D_rrho}, with the same reasoning as before one gets the following equation for 
$f(r)$:
\beq\label{rho ODE compact form}
\begin{aligned}
& \frac{2\lambda}{3}r\sqrt{1-K(r)r^2}f^\prime(r) + \left[\frac{8\lambda}{3}\sqrt{1-K(r)r^2} + 1\right]f(r) +\\ 
& -  2\left\{\frac{\left[r^3K(r)\right]^\prime}{3r^2}\right\}^\prime r^2_\mathrm{m} \sqrt{1-K(r)r^2} = 0,
\end{aligned}
\eeq
with the boundary condition $f(0)=0$. 

In Figure~\ref{fig:dpr_dr+drho_dr:f=R}, we show the pressure and energy density gradient profiles for positive and negative values of the anisotropy parameter $\lambda$. As one can clearly see, in the case where $\lambda<0$, there is a divergence of the pressure and energy density gradient profile in the center below a critical value. This behavior is due to the mathematical structure of \Eq{p_r ODE compact form} and \Eq{rho ODE compact form}, where the radial derivatives 
of $\tilde{p}_\mr$ and $\tilde{\rho}$ diverge at $r=0$, for $\lambda<-9/14$ and $\lambda<-3/8$, respectively. 

\begin{figure*}[ht!]
\centering
\includegraphics[width=0.497\textwidth]{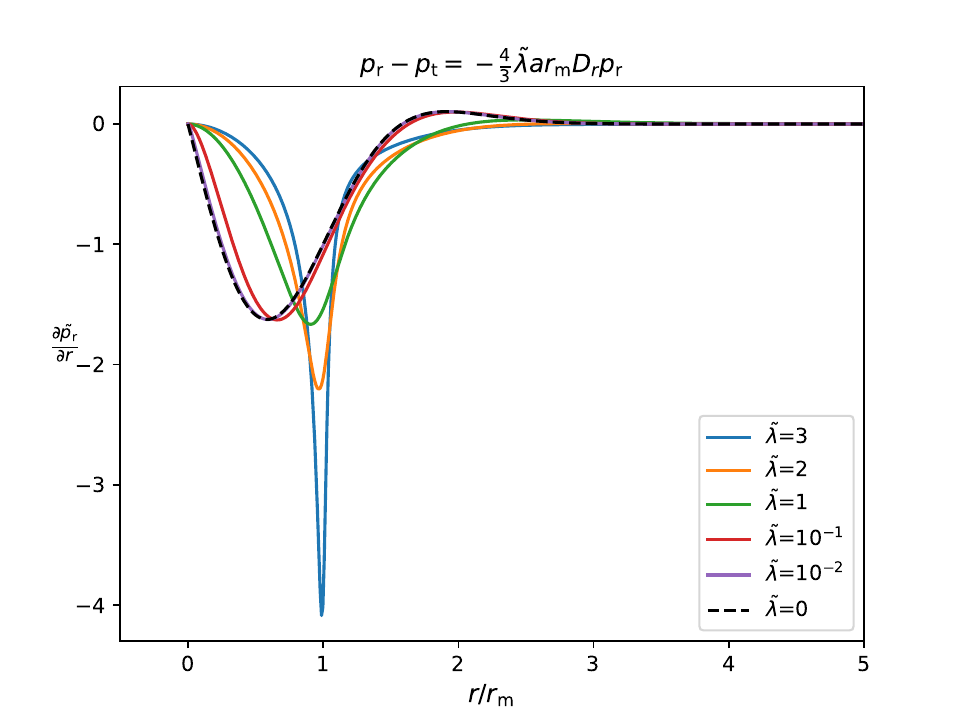}
\includegraphics[width=0.497\textwidth]{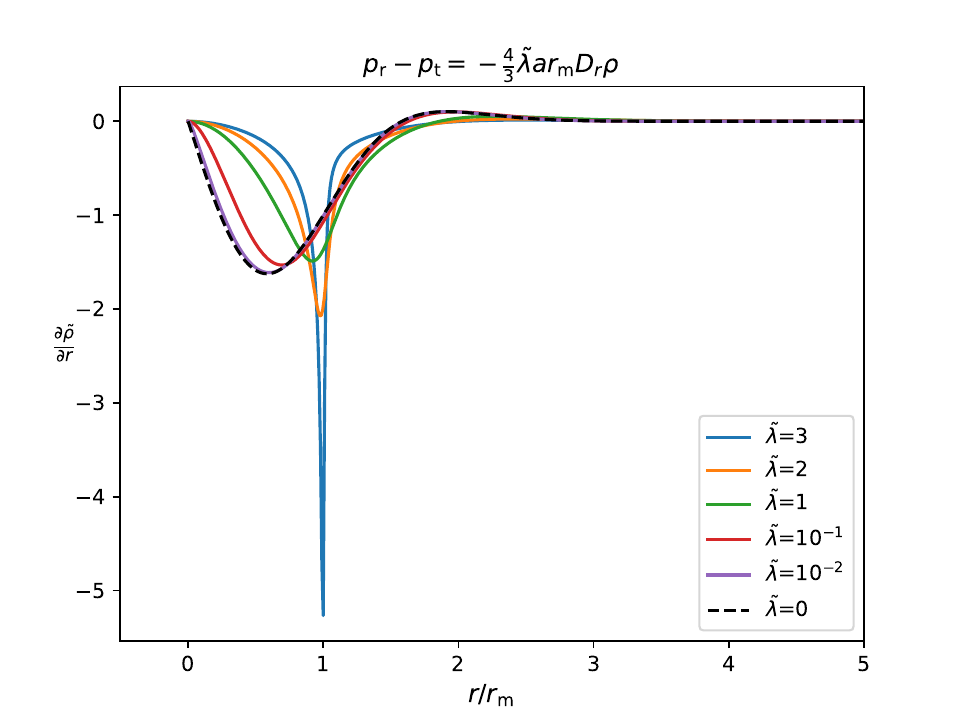}
\caption{In this figure we show the behavior of $\frac{\partial\tilde{p}_\mr}{\partial r}$ (left panel) and  $\frac{\partial\tilde{\rho}}{\partial r}$ (right panel) potted against 
$r/r_\mathrm{m}$ for $n=0$ and $\lambda>0$.}
\label{fig:dpr_dr+drho_dr_f=rho^n}
\end{figure*}

To see this more in detail, consider for example the EoS in terms of the pressure gradients (the same applies also for the energy density gradients) and develop $f(r)$ defined in \Eq{f_j:f=R} around zero as 
\[ f(r)=j_0 + j_1 r + j_2 r^2/2, \] 
where
\[ j_0 =f(0)= \tilde{p}^\prime_\mr(0) \,, \quad j_1=f^\prime(0) \quad \textrm{and} \quad j_2=f^{\prime\prime}(0) \]  
and then using the differential equation \eqref{p_r ODE compact form} we get that
\beq \label{rho ODE compact form around zero}
\begin{aligned}
& \frac{8\lambda r}{9}\left(1-\frac{\mathcal{A}r^2}{2}\right)\left(j_0 +j_1 r\right) \\ &  + \left[\frac{14\lambda}{9}\left(1-\frac{\mathcal{A}r^2}{2}\right)+1 \right]\left(j_0 +j_1 r + \frac{j_2 r^2}{2}\right) \\ & -\tilde{p}^\prime_\mathrm{r,iso}\left(1-\frac{\mathcal{A} r^2}{2}\right) = 0\,.
\end{aligned}
\eeq
Taking now the limit as $r\to0$ we obtain that
\[ \label{g_0}
j_0=  \tilde{p}^\prime_\mr(0) = \lim_{r\to 0} \frac{\tilde{p}^\prime_\mathrm{r,iso}(r)}{1+\frac{14\lambda}{9}},
\]
where $\tilde{p}^\prime_\mathrm{r,iso}=\frac{2}{3}\left\{\frac{\left[r^3K(r)\right]^\prime}{3r^2}\right\}^\prime r^2_\mathrm{m}$.

If $\lambda<-9/14$ one gets that $\tilde{p}^\prime_\mr(0) = 0^+$ which is not consistent because $\tilde{p}^\prime_\mr(0)$ should approach zero from negative values, 
namely $\tilde{p}^\prime_\mathrm{r,iso}(0)=0^{-}$. However, if $\lambda >-9/14$ one obtains the consistent result that $\tilde{p}^\prime_\mr(0) = 0^-$. For the critical value $\lambda=-9/14$, applying the De l'Hopital theorem and considering that $\tilde{p}^{\prime\prime}_\mathrm{r,iso}(0)<0$, one gets 
\[ 
\tilde{p}^\prime_\mr(0) = \lim_{r\to 0} \frac{\tilde{p}^\prime_\mathrm{r,iso}(r)}{1+\frac{14\lambda}{9}} = -\infty \neq 0^-\,.
\]
In the case of $p_\mathrm{r}-p_\mathrm{t} = -\lambda RD_\mathrm{r}p_\mr$ with $\lambda<0$ one gets that $\lambda$ should be 
larger than a critical value, namely \mbox{$\lambda>\lambda_\mathrm{c}=-9/14$}. When $p_\mathrm{r}-p_\mathrm{t} = -\lambda 
RD_\mathrm{r}\rho$, following the same procedure, one obtains that  \mbox{$\lambda>\lambda_\mathrm{c}=-3/8$} in order to avoid  
$\frac{\partial\tilde{\rho}}{\partial r}$ diverging at $r=0$.

\subsection{Equation of state with $g(r,t)=\rho^n(r,t)$}\label{A2}

In the case where the EoS is given in terms of pressure gradients, following \Eq{eq:p_r+p_t - f= rho^n-D_rp_r}, one should combine \Eq{p_tilde-rho_tilde - f=rho^n} 
and $\tilde{\rho}$ from \Eq{Perturbations-f=rho^n} to obtain after a straightforward calculation the following differential equation for the function $f(r)$:
\beq\label{p_r ODE compact form - f=rho^n}
\begin{aligned}
&\sqrt{1-K(r)r^2} \left[\frac{\rho^n_\mathrm{b}(a)}{\rho^n_\mathrm{b,i}}\frac{\chi_n(a)}{a} + \frac{1}{3}\right]\tilde{\lambda} rf^\prime(r)  \\ &  + \left\{\left[\frac{\rho^n_\mathrm{b}(a)}{\rho^n_\mathrm{b,i}}\frac{\chi_n(a)}{a} +\frac{4}{3}\right]\tilde{\lambda}\sqrt{1-K(r)r^2}+\frac{3r}{2r_\mathrm{m}}\right\}f(r)  \\ & -\left[\frac{\left(r^3K(r)\right)^\prime}{3r^2}\right]^\prime r_\mathrm{m} \sqrt{1-K(r)r^2} =0
\end{aligned},
\eeq
where $f(r)$ is defined in \Eq{f_j:f=rho^n} . The above differential equation should satisfy the boundary condition $\lim_{r\rightarrow 0}f(r)=0$ as imposed by \Eq{boundary condition}. 

Finally, in the case where the equation of state is given in terms of energy density gradients, following \Eq{eq:p_r+p_t - f= rho^n-D_rrho}, with the same reasoning as before one obtains the following differential equation for the function $f(r)$:
\beq\label{rho ODE compact form - f=rho^n}
\begin{aligned}
&\frac{2\tilde{\lambda}\sqrt{1-K(r)r^2}}{3} rf^\prime(r)  \\ & + \left\{\frac{8\tilde{\lambda}}{3}\sqrt{1-K(r)r^2}+\frac{r}{r_\mathrm{m}}\right\}f(r) \\ &  - 2 \left[\frac{\left(r^3K(r)\right)^\prime}{3r^2}\right]^\prime r_\mathrm{m} \sqrt{1-K(r)r^2} =0
\end{aligned},
\eeq

with $\lim_{r\rightarrow 0}f(r)=0$. 

In Figure~\ref{fig:dpr_dr+drho_dr_f=rho^n} we show the pressure and energy density gradient profiles for different values of  $\lambda$ 
and $n=0$. In this case, negative values of $\lambda$ lead to a divergence of the radial derivatives of $\tilde{p}_\mr$ and $\tilde{\rho}$ 
at $r=0$ and therefore they should not be taken into account. This can be seen by applying the same gradient expansion around zero
for \Eq{p_r ODE compact form - f=rho^n} as before, which gives 
\[ \label{h_0}
\tilde{p_\mr}^\prime(0) = \lim_{r\to 0} \frac{\tilde{p}^\prime_{\mathrm{r,iso}}(r)}{\frac{2}{3(1-2n)}\frac{\lambda\rho^n_\mathrm{b,i}}{\mu}\left(\frac{\mu}{\epsilon}\right)^3\left[(3-2n)\left(\frac{\mu}{\epsilon}\right)^{4n-2} -2\right]},
\]
where $\tilde{p}^\prime_\mathrm{r,iso}=\frac{2}{3}\left\{\frac{\left[r^3K(r)\right]^\prime}{3r^2}\right\}^\prime r^2_\mathrm{m}$,  and the 
necessary condition in order not to have a divergence at $r=0$ is 
\beq\label{lambda_c - f=rho^n - Drpr}
\frac{3}{2(1-2n)}\frac{\lambda}{\mu}\left(\frac{\mu}{\epsilon}\right)^3\left[(3-2n)\left(\frac{\mu}{\epsilon}\right)^{4n-2} -2\right]>0.
\eeq
From the above expression, fixing $\mu$ and $\epsilon$ one may identify two regimes, $n>1/2$ and $n<1/2$. In particular, when 
$n>1/2$ assuming that $\mu/\epsilon\ll 1$, one obtains that the second term within the square brackets of 
\Eq{lambda_c - f=rho^n - Drpr} is dominant and $\lambda>0$. On the other hand, if $n<1/2$ the first term within the brackets is now 
dominating, and one again gets $\lambda>0$. Therefore, if $\mu/\epsilon\ll 1$ one has in general that 
$\lambda>\lambda_\mathrm{c}=0$.

Finally, when the difference between the radial and the tangential pressure is proportional to the energy density gradients, by following the same reasoning, one gets the following necessary condition to avoid divergences around $r=0$
\beq\label{lambda_c - f=rho^n - Drrho}
\frac{4}{1-2n}\frac{\lambda}{\mu}\left(\frac{\mu}{\epsilon}\right)^3\left[ 1- \left(\frac{\mu}{\epsilon}\right)^{4n-2}\right]<0\,,
\eeq
and if $\mu/\epsilon_0\ll 1$, we again have $\lambda>0$ for any value of $n$.
\end{appendix}

\bibliography{PBH}
\end{document}